\documentclass[]{aa}
\usepackage{graphicx}
\usepackage{times}
\usepackage{natbib}
\bibpunct{(}{)}{;}{a}{}{,} 

\newcommand{\mpl}{M_{\mathrm p}}

\newcommand{\mearths}{$\mathrm{{M}_\oplus}$ }

\newcommand{\rp}{r_{\mathrm p}}
\newcommand{\pp}{\varphi_{\mathrm p}}
\newcommand{\be}{ \begin {equation}}
\newcommand{\ee}{ \end {equation}}
\newcommand{\op}{\Omega_\mathrm{p}}
\newcommand{\okep}{\Omega_\mathrm{K}}
\newcommand{\rg}{\mathrm{R_g}}
\newcommand{\cs}{c_\mathrm{s}}
\newcommand{\de}{D_\mathrm{e}}
\newcommand{\me}{\mathrm{M_\oplus}}
\newcommand{\phigpm}{\Phi_{\mathrm{Gp}m}}
\newcommand{\phigps}{\Phi_{\mathrm{Gps}}}
\newcommand{\phigp}{\Phi_{\mathrm{Gp}}}
\newcommand{\xs}{x_\mathrm{s}}

\newcommand{\eqref}[1]{(\ref{#1})}

\begin{document}
\title{On disc  protoplanet interactions
 in a  non-barotropic disc with thermal diffusion}
\author{Sijme-Jan Paardekooper \and John C. B. Papaloizou} 
\institute{DAMTP, University of Cambridge, Wilberforce Road, Cambridge
 CB3 0WA, UK \\ 
\email{S.Paardekooper@damtp.cam.ac.uk, J.C.B.Papaloizou@damtp.cam.ac.uk}} 

\date{Draft Version \today} 
  
\abstract{}{We study the disc planet interactions of low-mass
 protoplanets embedded in a circumstellar disc. We extend the standard
 theory of planet migration from the usual locally isothermal
 assumption to include non-barotropic effects, focusing on the
 validity of linear theory.}{We compared solutions of the linear
 equations with results from non-linear hydrodynamic simulations,
 where in both cases we adopted a background entropy gradient and
 solved the energy equation.}{We show that the migration behavior of
 embedded planets depends critically on the background  radial entropy
 gradient in the disc. The presence of such a  gradient not only
 changes the corotation torque on the planet, but also always
 guarantees a departure from linear behavior, which gives a  singular
 density response at corotation, in the absence of thermal or viscous
 diffusion. A negative entropy gradient tends to give rise to
 positive, non-linear corotation torques apparently produced  as
 material executes horseshoe turns at approximately constant
 entropy. These torques have no counterpart in linear theory, but can
 be strong enough to push the planet \emph{outwards} until saturation
 starts to occur after a horseshoe libration period. Increased thermal
 diffusion acts to reduce  these non-linear torques, but, at the same
 time, it can help to prevent their saturation. In combination with a
 small kinematic viscosity that is able to maintain a smooth density
 profile the positive torque could be sustained.}{} 
    

\authorrunning{S.-J. Paardekooper \& J.C.B. Papaloizou}
\titlerunning{Disc planet interactions in non-barotropic discs}

\maketitle


\section{Introduction}

During their formation process inside circumstellar discs, planets can
change their orbital parameters by gravitational interaction with the
gaseous disc \citep{1980ApJ...241..425G}. Torques generated at various
resonances can promote or damp eccentricity growth
\citep{2003ApJ...585.1024G}, and change the semi-major axis of the
forming planet \citep{1997Icar..126..261W}. One can distinguish
torques due to waves, generated at Lindblad resonances, which
propagate away from the planet, and corotation torques, generated near
the orbit of the planet, where material on average corotates with the
planet \citep[see][]{1979ApJ...233..857G}. The usual result of these
torques is that the planet migrates inwards, towards the central star,
at a rate that is determined by the planet mass and the disc
parameters \citep[see][]{1997Icar..126..261W}. 

In terms of planet and disc mass, three types of migration can be
distinguished. Low-mass planets, whose gravitational influence is not
strong enough to overcome pressure effects, generate a linear disc
response that gives rise to an inward migration rate that is
proportional to the planets mass. This is called Type I migration
\citep{1997Icar..126..261W}. For standard Solar nebula parameters,
Type I migration applies to planets up to several Earth masses
($\me$). 

Higher-mass planets excite non-linear waves, and are able to tidally
truncate the disc, opening an annular gap around their orbit
\citep{1979MNRAS...186..799L,1986ApJ...309..846L}. Migration occurring
when such a  gap is maintained is referred to as Type II migration
\citep{1997Icar..126..261W}, and it takes the planet inward on a
viscous time scale. The minimum mass for opening a gap, again for
standard disc parameters, is approximately one Jupiter mass.   

Both types of migration have been studied extensively through
numerical hydrodynamical simulations, in a two-dimensional set-up
\citep{2000MNRAS.318...18N,2002A&A...385..647D}  as well as in three
dimensions \citep{2003ApJ...586..540D,2003MNRAS.341..213B}, giving
good agreement in the respective mass regimes. In the intermediate
mass regime, depending on the disc viscosity and surface density
distribution, the corotation torque may  get a non-linear boost  to
the extent that it determines the sign of the total torque
\citep{2006ApJ...652..730M}. 

Corotation torques are also responsible for a third type of migration,
sometimes called runaway migration \citep{2003ApJ...588..494M}. This
Type III migration \citep{2004ASPC..324...39A} can be very fast, and
applies to planets that open up a partial gap and are embedded in a
massive disc \citep{2003ApJ...588..494M}.  
 
In all studies referred to above, a simplified equation of state,  for
which the pressure depends on the  density and the local radius  only
was used. The dependence on radius comes about from adopting a fixed
radial temperature profile. As at any location, the pressure depends
only on the density, we describe such an equation of state as being
locally barotropic.   The use of such an equation of state removes the
need to solve the energy equation and thus makes the equations more
tractable to tackle both analytically and numerically. The question
whether this approach is valid has only been addressed fairly recently
through numerical simulations that do include the energy equation
\citep{2003ApJ...599..548D, 2006A&A...445..747K}, focusing on
high-mass planets only. \cite{2003MNRAS.346..915M} studied the effect
of optically thin cooling on disc-planet interactions using a local
approach, while \cite{2005ApJ...619.1123J} calculated the torque on
low-mass planets analytically in discs with a realistic temperature
profile.   
 
 \cite{2006A&A...459L..17P}  showed that relaxing the barotropic assumption 
can change the migration behavior of low-mass planets dramatically. There it was shown,
 using three-dimensional radiation-hydrodynamical simulations, that, 
for the disc parameters adopted, low-mass planets  migrate \emph{outwards} 
  through the action of a strong positive corotation torque. Due to the  excessive
computational  overheads associated with  such simulations, it was not feasible to do a parameter 
 survey for different  planet masses and disc  models.

In this paper, we aim   to clarify the origin of the positive corotation torque,
 its dependence on planet mass and  disc parameters together with the
relationship of this type of disc planet interaction  to the  linear perturbation theory
 that has been successfully applied to the Type I migration regime.

 To ease the computational demands we focus on non-barotropic  two-dimensional discs
(commonly termed cylindrical discs, see \cite{2000ApJ...528..462H})  which incorporate thermal diffusion rather than the more complex  radiative transfer.

The plan of the paper is as follows. In Sect. \ref{secEq} we review the governing
 equations and the disc models used. Section \ref{secLin} is devoted to the linear theory of the interaction of a low-mass planet with a gaseous disc, and we solve the resulting equations in Sect. \ref{secLinCalc}.
 In order to remove singularities we adopted the well known Landau prescription
together with a small thermal diffusivity applied in some cases.
Although migration torques could be calculated in the limit of zero
dissipative effects, the density response became singular implying
that non-linear effects may always be important in  the corotation region.
 
After describing the numerical set-up in Sect. \ref{secSetup}, we perform fully non-linear hydrodynamical simulations in Sect. \ref{secNum}.
Indeed we find that  
a non linear small scale coorbital density
structure is set up at early times.  We go on to investigate the effect of thermal diffusivity,
finding that a non zero value is required to ensure the coorbital region is adequately resolved. 
The associated torque is then indeed found to be positive, for favorable density gradients, and sufficiently strong and negative entropy gradients, for about one horseshoe libration period after which saturation sets in. In Sect. \ref{secNonlin} we present a simple non-linear model of the corotation region based on material executing horseshoe turns with constant entropy, which can result in a density structure that leads to a temporary positive torque on the planet when there is a background negative entropy gradient and compare it with simulations. 

Whether the positive torque can be sustained for longer times will depend on the evolution of the planetary orbit and disc and whether these circumstances provide an appropriate corotational flow that can resupply appropriate low entropy material. To investigate this aspect, we study the long-term evolution of the torque in Sect. \ref{secLongterm}, and in the case of  a particular example,  show that for an appropriate  rate  of thermal diffusion and a small viscously driven mass flow through corotation, the entropy-related corotation torque can be sustained for several libration periods.
 
We go on to give a discussion of our findings in Sect. \ref{secDisc} and conclude in Sect. \ref{secCon}. 
 

\section{Basic equations and disc models}
\label{secEq}

The basic equations are those of the conservation of mass, momentum and energy for a two dimensional  cylindrical disc in a  frame rotating with angular velocity $\op$ in the form 
\be {\partial \rho \over \partial t}= -\nabla\cdot(\rho {\bf v}) \label{cont} \ee

\be {D {\bf v} \over D t} +2\op{\hat {\bf k}} \times {\bf v} = 
-{1\over \rho}\nabla P- \nabla\Phi \label{mot}\ee

\be \rho{ D E \over  Dt} - {P\over \rho} { D \rho \over D t}\equiv \rho T {D S \over  Dt}=
-\nabla\cdot{\bf F} .\label{energ}\ee
Here, $\rho$ denotes the  density, ${\bf v} = (v_r, v_{\varphi})$ the velocity, ${\hat {\bf k}}$ denotes the unit vector in the vertical direction, $P$ the  pressure and $\Phi =\Phi_\mathrm{G} +\op^2r^2/2 $
where $\Phi_\mathrm{G}$ is  the gravitational potential. The convective derivative is defined by 
\be {D\over Dt} \equiv  {\partial  \over \partial t}+ {\bf v}\cdot \nabla,\ee
and ${\bf F} = -K \nabla T$ is the heat flux, with $K$ being the thermal conductivity and $T$ being the temperature (of course radiation transport may be incorporated in this formalism). In our numerical simulations, we choose $K=K(r)$ such that the initial temperature profile gives $\nabla\cdot{\bf F} =0$. We adopt an ideal gas equation of state such that
\be P=\rg \rho T,\ee
with $\rg$ being the gas constant. The internal energy per unit   mass is given by 
\be E =\frac{P}{(\gamma-1)\rho}, \ee
where $\gamma$ is the constant  adiabatic exponent and \newline  $S=\rg \log (P/\rho^{\gamma})/(\gamma-1)$ is the entropy per unit mass.

We adopt a cylindrical polar coordinate system $(r,\varphi)$ with origin $(r=0)$  located at the central mass. The self-gravity of the disc is neglected. Thus the gravitational potential is assumed to be due to the central mass and perturbing planet such that $\Phi_\mathrm{G} = \Phi_\mathrm{G0}+ \Phi_\mathrm{Gp},$ where
\be \Phi_\mathrm{G0} =  -{GM_*\over r}, \ee
and
\begin{eqnarray}
\Phi_\mathrm{Gp} =  {-G\mpl\over\sqrt{r^2+\rp^2-2r\rp \cos(\varphi-\pp)+b^2 \rp^2}}
\nonumber \\
+{G\mpl r \cos(\varphi-\pp)\over \rp^2}.\label{pot}
\end{eqnarray}
In the above the last term is the indirect term, $M_*$ denotes the mass of the central object, with  $\mpl,$ $\rp$, and $\pp$ denoting  the mass, radial and angular coordinate  of the protoplanet respectively. We also allow for a gravitational softening parameter $b.$

We remark that there are several limiting cases in the above description. When heat transport is neglected, $(K=0)$,  the system is adiabatic. When the energy equation is dropped and the temperature is taken to be a fixed function of $r$ we have the usual locally isothermal limit. When $\op=0$ the reference frame is non-rotating but non-inertial as the origin accelerates together with the central mass. This is
accounted for by the indirect term in the potential. Numerical calculations are most conveniently 
performed in a frame corotating with the protoplanet. Then $\op$ becomes the circular Keplerian angular velocity at radius $\rp$. At a general radius, $r$, this is given by $\okep = (GM_*/r^3)^{1/2}$.

\begin{figure*}
\begin{center}
\includegraphics[width= \textwidth]{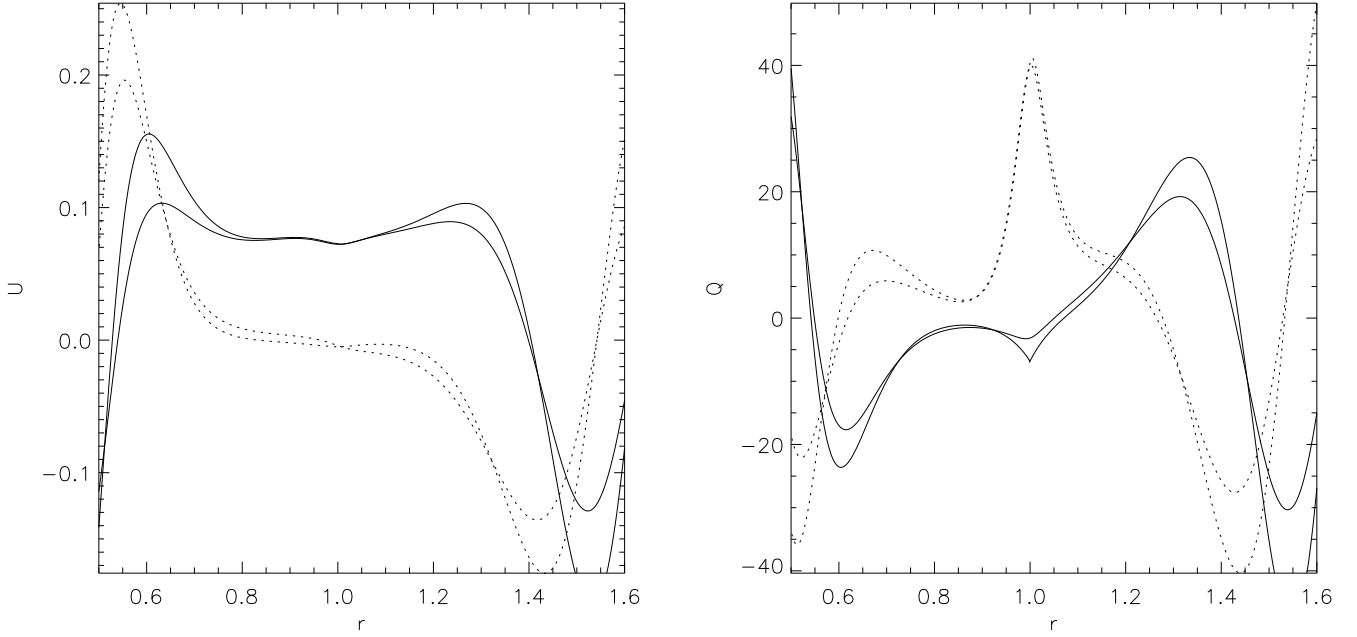}
\caption{The  functions $U$ and $Q$ divided by the protoplanet to central star mass ratio are plotted
for $\rho_0 \propto r^{-1/2},$ $b=0.03$  and $m=2$ in dimensionless units. Real parts are plotted with
dashed lines , imaginary parts with  full lines. Curves for the Landau parameters $\epsilon= 10^{-5}$
$\epsilon= 3.\times10^{-2}$ are shown. The response is very similar for these cases but the  latter case is slightly  more damped which is manifest by somewhat smaller amplitude swings.
}
\label{fig00}
\end{center}
\end{figure*}

The formulation given above applies to a cylindrical disc model 
where there is no vertical stratification or dependence on 
the vertical coordinate. 
The equations  apply to a cylindrical  system
 with  a constant vertical semi-thickness, $H_0$
which may be specified arbitrarily as there is no explicit dependence
on it.
The  density, $\rho,$ is taken to have the same  spatial  variation as the
surface density to which it is related through $\Sigma =2\rho H_0.$
 For a thin disc of the type we  wish to consider,
 at any location there is a putative local vertical semi-thickness 
$H = \cs/\okep$, with $\cs =\sqrt{P/\rho}$ being the local 
isothermal sound speed that is  a function of position and 
associated with the neglected vertical stratification. 
When considering the physical state variables associated
with any such location we regard $H_0$ as having been adjusted to coincide with $H.$

\subsection{Equilibrium models}
In this paper we work with  equilibrium model discs  into which a perturbing  protoplanet is subsequently inserted. These are such that density has a power law dependence of the form $\rho \propto  r^{-\alpha}$, with $\alpha$ being constant. We also adopt $T \propto r^{-1}$, giving a putative semi-thickness $H \propto r$, then we have $P \propto r^{-\alpha-1}$. The local  gas angular velocity required for radial hydrostatic equilibrium is given by $\Omega = \okep (1-(\alpha +1)(H/r)^2)$.
Thus for these models $\Omega \propto \okep$ and the epicyclic frequency $\kappa =\Omega$. Here we recall in general that $\kappa^2 = (2\Omega/r) d(r^2\Omega)/dr$. We also adopt the representative value  $\gamma =1.4$ unless stated otherwise.


\section{Linear theory}
\label{secLin}

We here consider the response of the disc to the forcing of a protoplanet in a circular orbit assuming that the induced perturbations are small so that the basic equations may be linearized. We perform a Fourier decomposition of the perturbing potential such that in the non-rotating frame
\be \Phi_\mathrm{Gp} = {\cal R}_e \left(\sum_{m=0}^{\infty}
 \phigpm\exp{(im\varphi-im\okep t)}\right),\ee 
where ${\cal R}_e$ denotes that the real part is to be taken. Here, as well as in the rest of this section, $\okep$ is short for $\okep(\rp)$.

Similarly, the perturbation response of state variables that are non-zero in the equilibrium, indicated with a prime, for each value of $m$, has an exponential factor $\exp{(im\varphi-im\okep t)}$ with an amplitude depending only on $r$. Equations for these amplitudes are obtained by linearizing the basic equations. We begin by considering the adiabatic limit in which the thermal diffusivity is set to zero. Then we linearize the adiabatic condition
\be {DS \over Dt} = {\partial S \over \partial t}+{\bf v}\cdot \nabla S =0 ,\ee
to obtain
\be  S' = -{v_r\over i{\overline \sigma}} {d S \over dr},  \label{lent}\ee
where ${\overline \sigma} = m(\Omega-\okep).$
Expressed in terms of the density and pressure perturbations Eq. (\ref{lent}) takes the equivalent form
\be  {P'\over \gamma P} -{ \rho'\over \rho}= 
-{v_r{\cal A}\over i{\overline \sigma}g_r}\label{pent},\ee
where $g_r = -(1/\rho)(dP/dr)$ 
and the square of the Brunt-V\"ais\"al\"a frequency is given by
\be {\cal A} = -{1\over \rho} {d P \over dr} \left({1 \over \gamma P}{d P \over dr}
 -{1\over \rho}{d \rho \over dr}\right),\label{BVfrq}\ee
which can be negative in regions of formal local convective instability. Equation (\ref{pent}) can be used to express the density perturbation in terms of the pressure and velocity perturbations.

\begin{figure*}
\begin{center}
\includegraphics[width= \textwidth]{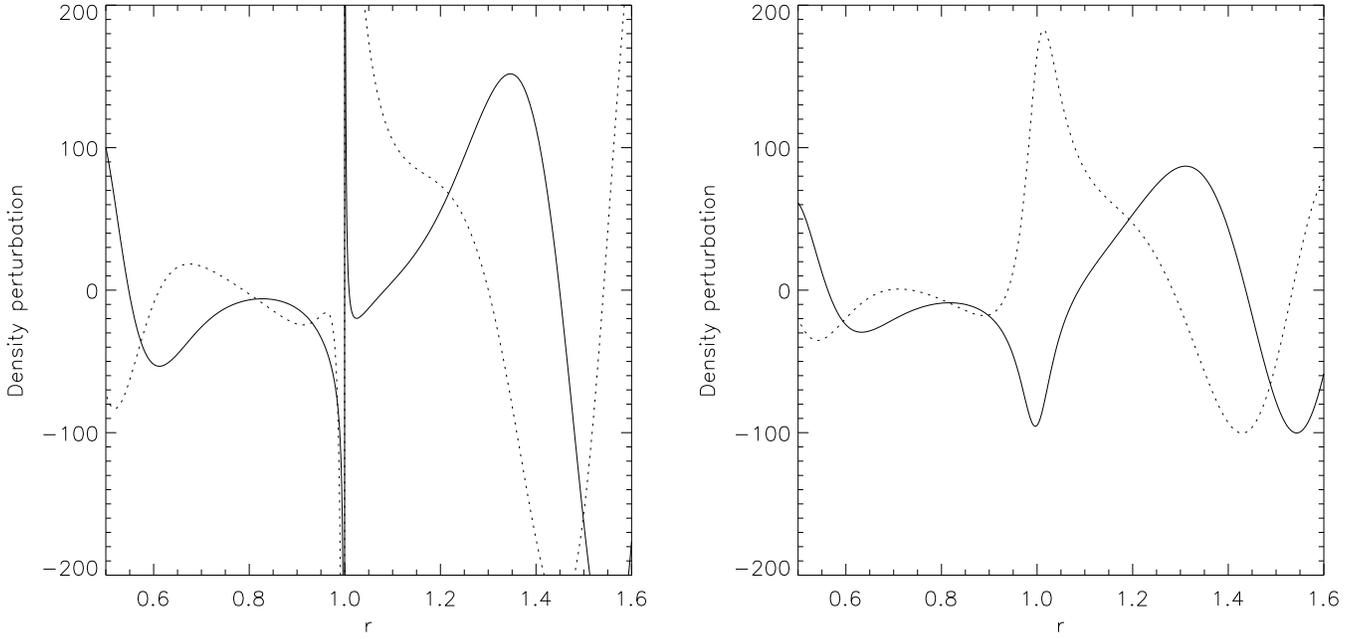}
\caption{The  density response divided by the  protoplanet
 to  central star mass ratio,
for $\rho_0 \propto r^{1/2},$ $b=0.03$  and $m=2$
  in dimensionless units. Real part dashed line , imaginary part full line.
For the left panel the Landau parameter  $\epsilon= 10^{-5}$ and  for the
 right panel
$\epsilon= 6\times10^{-2}.$
In the former case the amplitudes of the real and imaginary parts
reach extreme values of order $10^5$. }
\label{fig01}
\end{center}
\end{figure*}

\begin{figure*}
\begin{center}
\includegraphics[width= \textwidth]{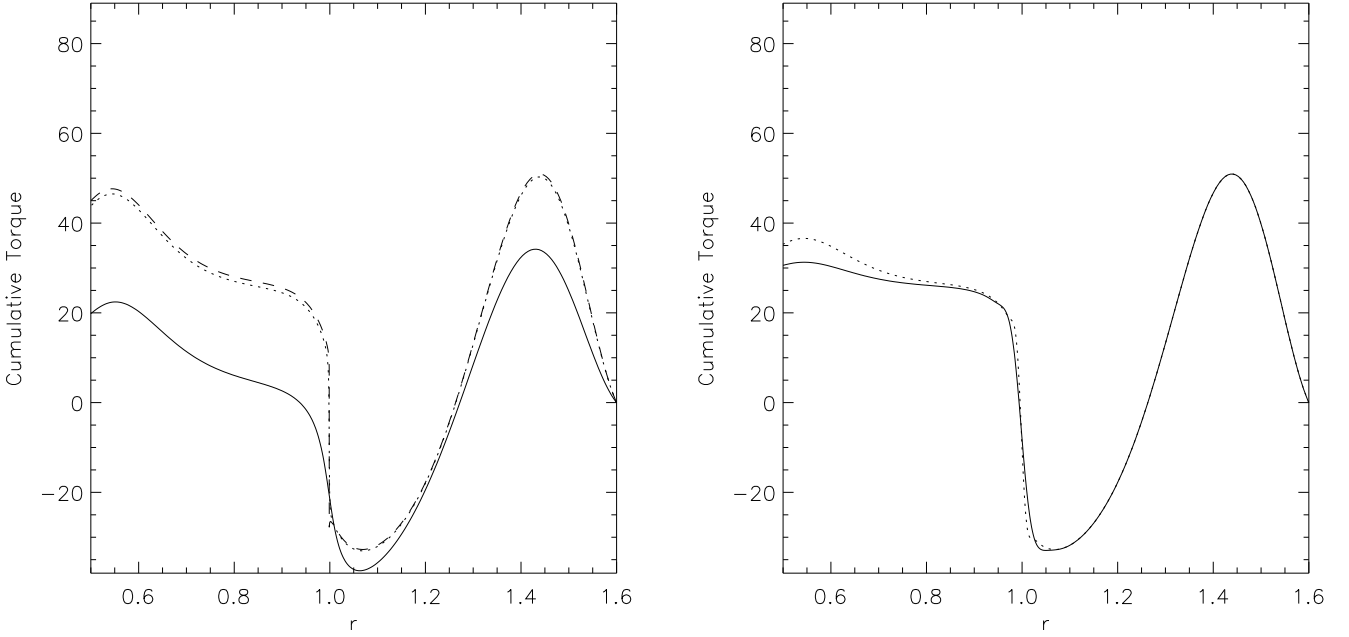}
\caption{The  cumulative torque in units of $ q^2 \Sigma(\rp)\okep^2\rp^4$ acting on the protoplanet plotted as a function of dimensionless radius for $\rho_0 \propto r^{-1/2},$ $b=0.03$  and $m=2.$ For the left panel plots are given for the Landau parameters  $\epsilon= 10^{-5}$ (dashed line)
$\epsilon= 10^{-4}$ (dotted line) $\epsilon= 3\times 10^{-2}$ (full line). For the right panel plots are given for the dimensionless thermal diffusivities $\de = 10^{-6}$ (dotted line) and  $\de =  10^{-5}$ (full line). The cumulative torque is defined to be zero at the outer boundary and equal the total torque at the inner boundary. In all of these cases the total torque is positive. }
\label{fig02}
\end{center}
\end{figure*}

Linearization of the two components of the equation of motion together with the continuity equation then yield, after elimination of the azimuthal velocity perturbation, a pair of first order ordinary differential equations for the quantities $Q= P'/P^{1/\gamma}$ and  $U =r P^{1/\gamma} v_r$ which may be written in the form   
\be {dQ\over dr} = C_1v_r +C_2Q + S_1, \label{DQ} \ee
\be  {dU \over  dr} = D_1Q +  D_2v_r  + S_2, \label{DU}\ee
where  the coefficients are given by 
\begin{eqnarray}
C_1 = -{i\rho P^{-(1/\gamma)}}
({\overline \sigma} -(\kappa^2 + {\cal A})/{\overline \sigma}) ,\nonumber \\
C_2 = -(2 m \Omega)/(r {\overline \sigma}),  \nonumber \\
D_1 = -i r({\overline \sigma}^2 - \gamma P m^2/(r^2\rho))/
({\overline \sigma}\gamma P^{1-2/\gamma}), \nonumber \\
D_2 = (P^{1/\gamma} m \kappa^2)/(2 \Omega {\overline \sigma}),\nonumber \\
S_1 = -{\rho P^{(-1/\gamma)}}[(d \phigpm/dr) +
 (2 m\Omega)\phigpm/(r {\overline \sigma})],  \nonumber \\
S_2 = i( P^{1/\gamma} m^2)/(r {\overline \sigma})  \phigpm. \nonumber
\end{eqnarray}

\subsection {The corotation singularity}
The perturbation of the protoplanet causes the excitation of outgoing density waves that are associated with a conserved wave action or angular momentum flux. This causes a torque to act on the protoplanet.
However, angular momentum exchange between protoplanet and disc may also occur directly at corotation where ${\overline \sigma}=0.$ In linear theory, this type of  exchange is localized at corotation through the operation of a  corotation singularity. In order to study the domain near corotation where ${\overline \sigma}=0,$ we neglect ${\overline \sigma}^2$ in the first set of brackets in the expression 
for $D_1$ above. Then we may derive a single equation for $U$ from Eqs.  (\ref{DQ}) and (\ref{DU}) in the form
\begin{eqnarray} 
{ P^{2/\gamma}\over \rho r} {d\over dr}\left( {\rho r\over P^{2/\gamma}}
\left ( {d U \over dr}\right)\right)= \nonumber \\
U\left[ {m^2\over r^2}\left(1-{{\cal A}\over{\overline \sigma}^2}\right)
+ { m P^{2/\gamma}\over \rho r {\overline \sigma}}
{d\over dr}\left( \rho\kappa^2 \over 2\Omega P^{2/\gamma}\right)\right] \nonumber \\
-{i\phigpm m^2 {\cal A} P^{1/\gamma}\over {\overline \sigma}rg_r}. \label{cosing}
\end{eqnarray}

We see that Eq. \eqref{cosing} is in general singular at corotation with a second order pole at  ${\overline \sigma}=0$. This will result in angular momentum exchange with the perturber. In order to be singularity free we require both that the entropy gradient be zero or equivalently ${\cal A}=0$, and 
that $d(\rho\kappa^2/ (2\Omega P^{2/\gamma})/dr =0$. The latter condition is the generalization of the condition that the gradient of specific vorticity, or vortensity, given by
\begin{equation}
\frac{d}{dr}\left(\frac{\kappa^2}{2\Omega\rho}\right),
\label{eqdefvort}
\end{equation}
should  be zero that applies in the strictly barotropic case.

\begin{figure*}
\begin{center}
\includegraphics[width= \textwidth]{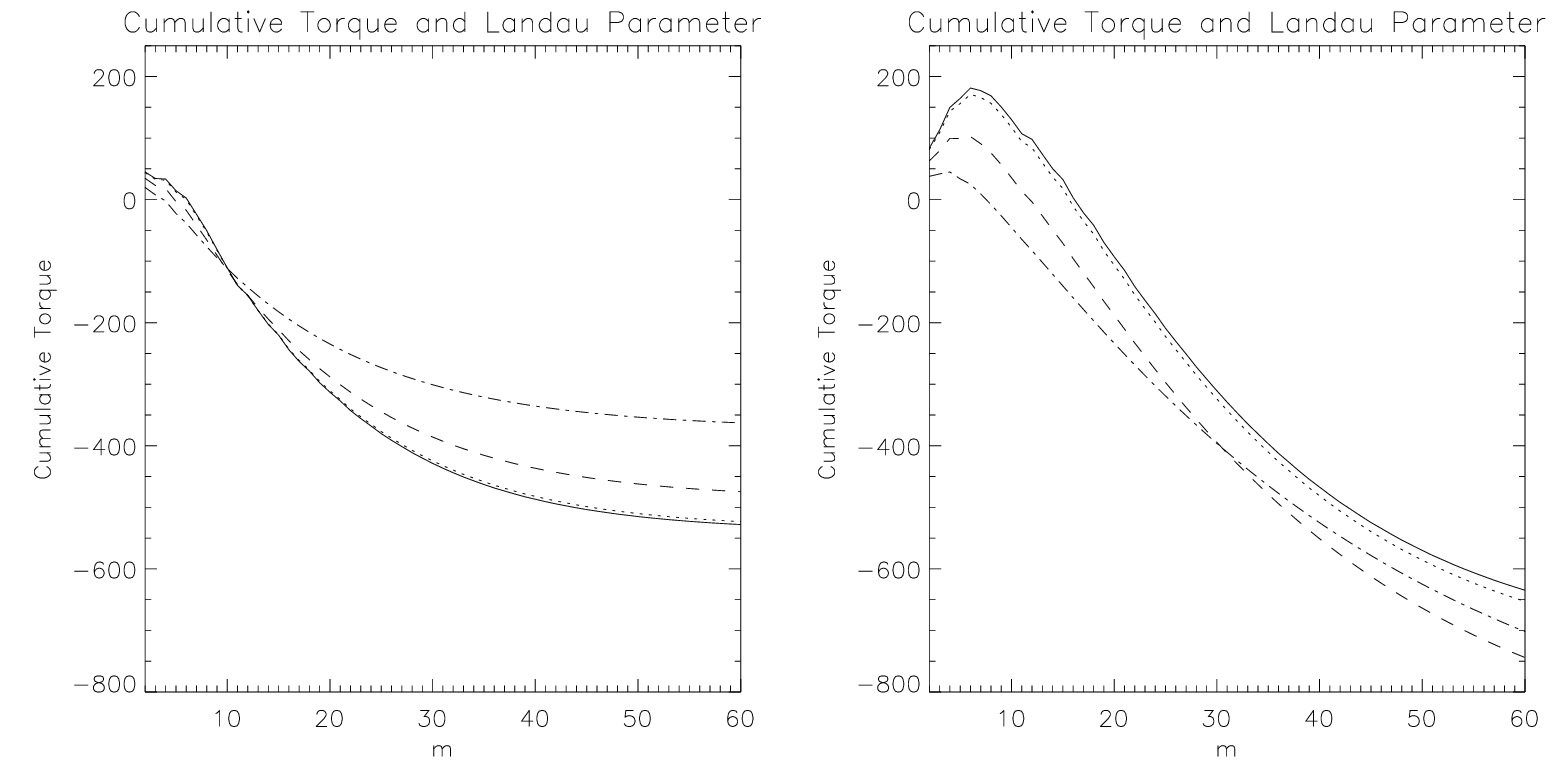}
\caption{The  cumulative torque in units of $q^2 \Sigma(\rp)\okep^2\rp^4$ acting on the protoplanet  as a function of $m$ for $\rho_0 \propto r^{-1/2},$ $b=0.03$ (left panel) and $b=0.01$ (right panel). For  both panels plots are given for the Landau parameters  $\epsilon= 10^{-5}$ ( full line), $\epsilon= 10^{-4}$ (dotted line), $\epsilon= 10^{-3}$ (dashed line), $\epsilon= 3\times 10^{-2}$ (dot dashed line).}
\label{fig03}
\end{center}
\end{figure*}

\subsection{The Landau prescription}
In order to solve the forced response problem, we have to deal with the corotation singularity. Physically this is resolved by including the dissipative effects of heat transport and viscosity. However, by increasing the order of the system of equations, this significantly increases the complexity of the problem. For many purposes this can be avoided by using the Landau prescription. According to this
the forcing frequency is given a small positive imaginary part such that ${\overline \sigma} \rightarrow m(\Omega-\okep-i\epsilon\okep),$ where $\epsilon$ is very small. In practice, as is the case with the problem considered here, the torques between  disc and protoplanet converge as $\epsilon \rightarrow 0$ even though the solution becomes singular.

The latter feature means that even though torques may be convergent the response becomes non-linear near corotation making the linear theory invalid. The time $1/(\okep\epsilon)$ may be interpreted as the time scale over which the perturbing potential is turned on. Thus a non-linear breakdown for $\epsilon$ below a certain value, $\epsilon_0,$ means that, viewed from the point of view of an initial value problem, the linear solution should be considered as being valid only for a finite time period, which may be estimated as to be of order $1/(\okep\epsilon_0)$. 

For the problem on hand, non-linear breakdown occurs mainly through terms associated with the entropy gradient or ${\cal A}$, these being associated with a second order pole. This type of singularity may be removed by introducing heat diffusion so that we would expect that for a sufficiently high thermal diffusivity, the validity of linear theory should be restored. This we explore below.

\subsection{The effect of a small thermal diffusivity}\label{LinDe}
In order to investigate the effect of thermal diffusion on the corotation singularity, we now consider the modification of the condition given by Eq. \eqref{lent} that occurs when a small thermal diffusivity is present.

In this case we must linearize the full energy equation, Eq. \eqref{energ}, which leads to
\be  i{\overline \sigma}S' + v_r {d S \over dr}= {K\over \rho T}\nabla^2 T'.\ee
Here, because of rapid variation near corotation, when considering terms involving $K,$ we neglect  everything other than the second derivative term.  Assuming that the variation of the pressure perturbation can be neglected, (this approach can be validated by inspection of the form of the solutions), we  then have $S' = (\partial S/\partial T)_p T',$ and accordingly  
\be i{\overline \sigma}S' + v_r {d S \over dr}= {K\over C_p \rho }\nabla^2 S', \label{lentD} \ee
where $C_p =\gamma \rg/(\gamma-1)$ is the specific heat at constant pressure. As we are interested in a local region around corotation, we set $r=\rp+x,$ where $x \ll \rp,$ and    ${\overline \sigma}\rightarrow
-3m \okep x/(2\rp).$ In addition we replace $\nabla^2 \rightarrow (d^2/dx^2 -m^2/\rp^2).$

Then Eq. \eqref{lentD} can be written in the form
\be {d^2 S'\over dz^2} -i\beta z S' = F,\label{rent} \ee
where $z= x+2Dmi/(3\rp\okep),$ $\beta = -3m\okep/(2\rp D),$ $D=K/(\rho C_p),$ and $F =  v_r (d S / dr)/D,$ the last three quantities being assumed constant. Equation \eqref{rent} can be solved in terms of in-homogeneous Airy functions \citep[e.g..][]{1970hmf..book.....A} in the form
\be S'=-\left({3^{1/3}\pi F\over (-\beta)^{2/3}}\right) Hi(\zeta)\label{Dsol}\ee
with 
\be \zeta = -{2(i{\overline \sigma}+\epsilon_1) (9m/2\de)^{1/3}\over 3m \okep}\ee
and
 \be \epsilon_1  = m^{2} \de \okep,\ee
with $\de$ being   the  dimensionless diffusivity $\de = D/(\rp^{2} \okep).$ Here all quantities are evaluated at corotation, $r=\rp,$ apart from ${\overline \sigma} \rightarrow -3m \okep x/(2\rp).$ 

The in-homogeneous Airy function   $Hi(\zeta)$ \citep[see][]{1970hmf..book.....A} is defined by
 \be Hi(\zeta) ={1\over \pi}  \int_0^{\infty} \exp{(-k^3 +k\zeta)} dk.\ee
Using this solution, $S'$ is no longer singular  at corotation. By comparing Eq. \eqref{lent} for $S'$ with the solution determined by Eq. \eqref{Dsol} we can  see how to remove the corotation singularity.

It is apparent that, {\it where it appears multiplied by the entropy gradient,} the singular quantity
\be {\Omega\over \overline \sigma} \rightarrow
{2\pi i\over 3m}\left({9m\over 2\de}\right)^{1/3}
Hi\left({-2(i{\overline \sigma}+\epsilon_1)(9m/2\de)
^{1/3}\over 3m \okep}\right).\ee
In this way the divergence at $\overline \sigma=0$ associated with  terms that involve the entropy
gradient is removed. When we implemented cases with thermal diffusivity as described below, wherever $\overline \sigma$ otherwise occurred, we applied the Landau prescription with $\epsilon = 10^{-5}.$

From Eq. \eqref{rent}, the length scale associated with the diffusively controlled corotation region is $|\beta|^{-1/3} \sim \rp(\de/m)^{1/3}.$ Thus for sufficiently large $\de,$ we expect that the effects of corotation associated with a radial entropy gradient should be reduced such that the validity of linear theory is restored as far as these are concerned. Additional restrictions may result from a radial vortensity gradient.


\section{Linear response calculations}
\label{secLinCalc}

We have solved Eqs. \eqref{DQ} and \eqref{DU} for a variety of disc models, softening parameters  and values of $m.$ We have used  both the Landau prescription and the combination of that, together with a finite thermal diffusivity used to deal with terms associated with the entropy gradient, to deal with the corotation singularity. The equations were integrated using a fifth order Runge-Kutta scheme
and outgoing wave conditions were applied. In practice these were applied at radii that became  closer to the protoplanet as $m$ increased.

For convenience throughout, we adopt a system of units for which $\rp=1,$  $\okep(\rp) =1,$  and the density of the unperturbed disc at the protoplanet orbital radius $\rho(\rp)=1.$ The orbital period of 
the protoplanet is then $2\pi.$ In all cases the aspect ratio $\cs/(r\Omega)$ was taken to be $0.05.$  
 
\begin{figure*}
\begin{center}
\includegraphics[width= \textwidth]{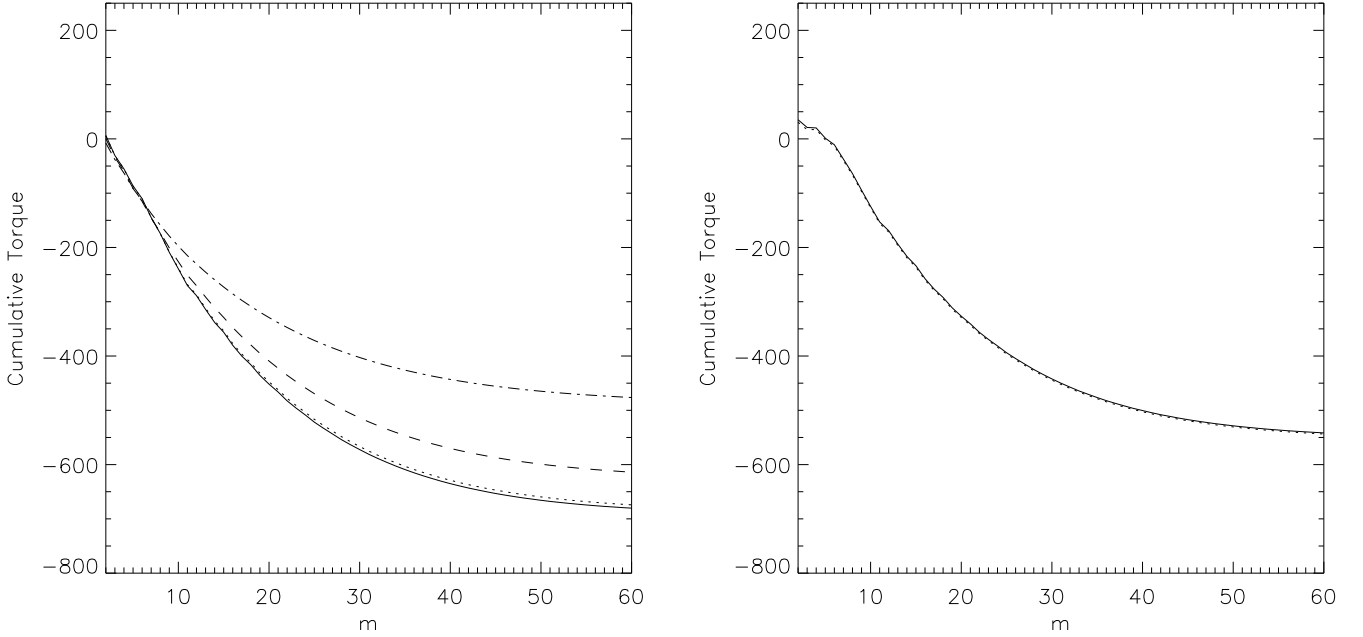}
\caption{The  cumulative torque in units of $q^2 \Sigma(\rp)\okep^2\rp^4$ acting on the protoplanet  as a function of $m$ for $\rho_0 \propto r^{-3/2}$ and  $b=0.03$ (left panel) and for  $\rho_0 \propto r^{-1/2},$ $b=0.03$ but with thermal diffusion  (right panel). For  the   left panel plots are given for the Landau parameters  $\epsilon= 10^{-5}$ ( full line), $\epsilon= 10^{-4}$ (dotted line), $\epsilon= 10^{-3}$ (dashed line), $\epsilon= 3\times 10^{-2}$ (dot dashed line). In  the right panel plots are given for 
the dimensionless thermal diffusivities $\de = 10^{-6}$ (dotted line) and  $\de =  10^{-5}$ ( full line).
The curves almost coincide for these cases.}
\label{fig04}
\end{center}
\end{figure*}

\begin{figure}
\centering
\resizebox{\hsize}{!}{\includegraphics[]{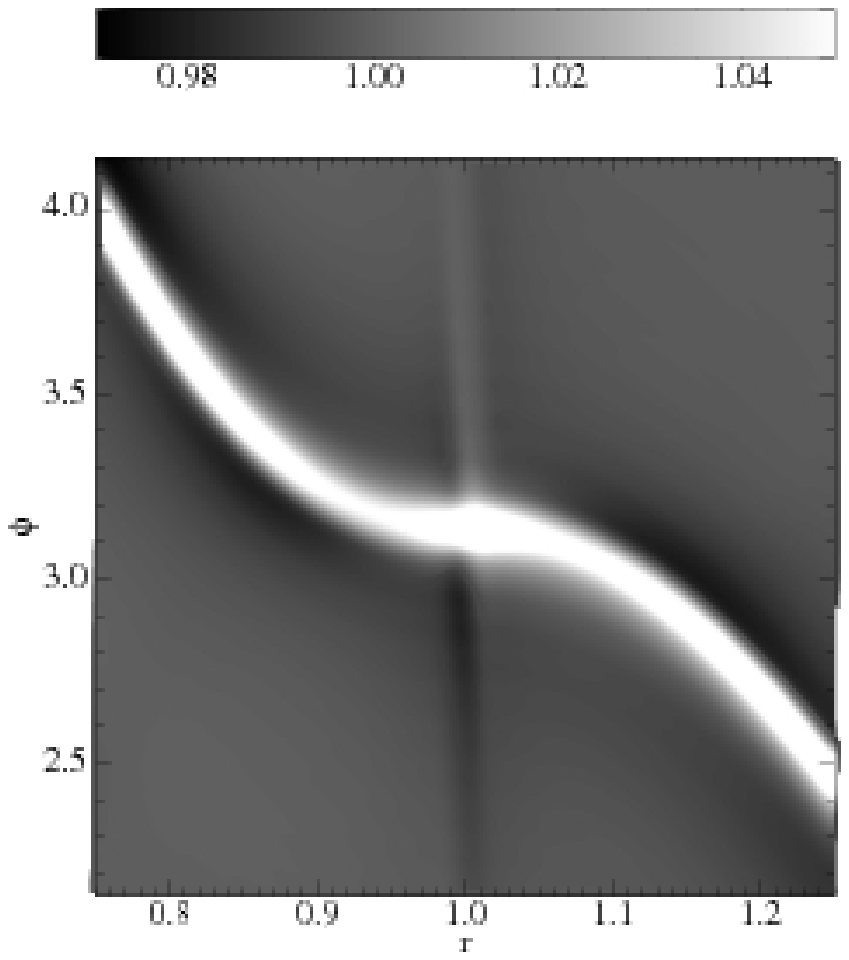}}
\caption{Contour plot of the full linear  density response for $\rho_0 \propto r^{-1/2}$ and thermal diffusivity $\de=10^{-6}$ in dimensionless units. The linear theory is only likely to  be valid for planet masses significantly 
less than an earth mass in this case (see Eq. \eqref{cond1}). Note the  narrow high and low density regions on the corotation circle near the protoplanet. These, also seen in the non-linear simulations,  respectively lead and trail the protolanet giving rise to a positive corotation torque. However, the effects of this torque are not strong enough to counteract the negative Lindblad torques contributed by the high density wakes.}
\label{fig0}
\end{figure}

\subsection{Results}
The  functions $U$ and $Q$ are  plotted for the disc with $\rho_0 \propto r^{-1/2},$ $b=0.03$, being comparable to $H/r$, 
and $m=2$ in Fig. \ref{fig00}. In these cases a Landau prescription was used to deal with  the
corotation singularity. We show curves for the Landau parameters $\epsilon= 10^{-5}$ and $\epsilon= 3\times10^{-2}.$ As is indicated in Fig. \ref{fig00}, the response functions are very similar in these cases for which the Landau parameter varies by a factor of $3000.$ The  $\epsilon= 3.\times10^{-2}$ case not unexpectedly  shows slightly  more damping. 

However,  the density response exhibits different behavior. This is plotted  for the Landau parameters  $\epsilon= 10^{-5}$ and  $\epsilon= 6\times 10^{-2}$ in Fig. \ref{fig01}, for a disc with $\alpha=-1/2$ (very similar behavior occurs for $\alpha=1/2$). This response shows a strong singularity as $\epsilon$ is decreased to zero. The extreme values  reached  are $\propto 1/\epsilon.$
As the singularity is approached the radial width $\propto \epsilon.$ The effect of this is that although the pressure and velocity response as well as the migration torque converge as $\epsilon \rightarrow 0,$
the density response becomes increasingly singular. This means that linear theory always breaks down in that limit. This breakdown is in fact associated with terms $\propto {\cal A},$ the entropy gradient and it can be removed by considering thermal diffusion. To deal with this, one can use the relative smoothness of the pressure and velocity perturbations to  obtain a simple description governed by a second order ordinary differential equation (see above). In order to remain in a regime for which linear theory is valid as far as these are concerned it is essential to either incorporate a non zero thermal diffusivity or restrict consideration to values of $\epsilon$ that are high enough for linear theory to be valid. The latter viewpoint can be regarded as restricting consideration of evolution times for an initial value problem to values smaller than $ \sim 1/(\epsilon\okep(\rp)).$  

\subsection{Migration torque}
The torque acting on the protoplanet is calculated directly by summing the  torque integrals evaluated for each Fourier component in the form
\be {\cal T}(r_\mathrm{in}) = 2H{\cal I}_m\left[\int^{r_\mathrm{out}}_{r_\mathrm{in}}\pi m r \rho' \phigpm dr\right],\label{eqmigtorque}\ee
where the integral is taken over the unperturbed disc domain $(r_\mathrm{in},r_\mathrm{out})$ and
${\cal I}_m$ denotes that the imaginary part is to be taken. The factor $2H$ takes account of the vertical direction by assuming the cylindrical disc model to apply over a vertical extent $2H.$ We  use  the above to  define a cumulative torque as a function of $r$ for each $m.$  
Thus 
\be {\cal T}(r) = \int^{r_\mathrm{out}}_{r} T_\mathrm{p}(r)dr,\ee
where $ T_\mathrm{p}(r)$ is the torque density. Then the total torque acting on the protoplanet is ${\cal T}(r_\mathrm{in}).$

The  cumulative torque acting on the protoplanet is plotted as a function of dimensionless radius
for $\rho_0 \propto r^{-1/2},$ $b=0.03$  and $m=2$ in Fig. \ref{fig02}. Plots are given for cases using the Landau prescription with parameters  $\epsilon= 10^{-5},$ $\epsilon= 10^{-4}$ and  $\epsilon= 3\times 10^{-2}.$ For comparison plots are given for the dimensionless thermal diffusivities $\de = 10^{-6}$ (dotted line) and  $\de =  10^{-5}$ implemented as described above.  In all of these cases the net torque is positive. We remark that the curves for $\epsilon= 10^{-5}$ and $\epsilon= 10^{-4}$ are very close as are those for $\de = 10^{-6}$  and  $\de =  10^{-5}.$ This indicates good convergence of the torques as dissipative effects are reduced.

To emphasize this feature we show the  cumulative torque acting on the protoplanet  as a function
of $m$ for $\rho_0 \propto r^{-1/2}$ for the two softening parameters $b=0.03$  and $b=0.01$ in Fig. \ref{fig03}. The softening parameter that should be used is uncertain but it should be of the order of the aspect ratio to simulate 3D effects. Plots are given for the Landau parameters  $\epsilon= 10^{-5},$ 
$\epsilon= 10^{-4},$ $\epsilon= 10^{-3}$ and $\epsilon= 3\times 10^{-2}.$ The total torques are always negative corresponding to inward migration. Again the convergence for small $\epsilon$ is good but note that torques of larger magnitude are obtained for the smaller softening parameter and larger $m$ need to be considered.

Similar results are obtained for different disc models and when thermal diffusion is employed as illustrated in Fig. \ref{fig04}. There we plot cumulative torques as a function of $m$ for $\rho_0 \propto r^{-3/2}$ and  $b=0.03.$ We show also the corresponding plots for  $\rho_0 \propto r^{-1/2},$ $b=0.03$ but with the implementation of thermal diffusion.

As indicated above, linear theory always breaks down for sufficiently small dissipation parameters, and it is important to estimate parameter regimes where it could be valid. We have done this both when a pure Landau prescription is used and also when thermal diffusion is applied. To do this we calculate the full
response by summing over $m$ and determine the condition that the perturbation to the entropy gradient be less than or equal to the same magnitude as the unperturbed value. 
Clearly non-linear effects may set in for weaker perturbations
than those that result in this condition being marginal.
Thus we estimate that non-linearity sets in for
smaller diffusion coefficients than those obtained when this condition
is marginally satisfied. We show a contour plot of the  full linear density response for $\rho_0 \propto r^{-1/2}$ and thermal diffusivity $\de=10^{-6}$ in Fig. \ref{fig0}. From this and other similar responses we obtain the condition
\be \left({0.03\over b}\right)^{1/3}
   {(1.55\times (q/10^{-5}))
   \over {\left(\de/10^{-6}\right)^{2/3}}} < 1.\label{cond1}\ee
Similarly for a pure Landau prescription, we obtain
\be \left({0.03\over b}\right)^{1/4}
{2.5\times 10^{7}q \over \left(\epsilon/10^{-3}\right)^{2}} < 1.\label{cond2}\ee
Interestingly, from Eq. \eqref{cond1} we estimate that for protoplanets in the earth mass range, if  $\de < 10^{-6},$ there should be departures from linear theory. Similarly from Eq. \eqref{cond2}, we estimate  
protoplanets in this mass range to be in the non-linear regime if $\epsilon <\sim  10^{-2}.$ 
This  indicates that linear theory may only be relevant for short 
evolutionary times in cases with very low dissipation  and then from Fig. \ref{fig03} we would expect that the
full linear torque is never established.
 Although the above estimates are uncertain, we comment that the scaling implied by Eq. (\ref{cond2}), that for the same degree of nonlinearity $\epsilon \propto q^{1/2},$ implies that the time required to attain the same degree of departure from linear theory should scale as $q^{-1/2}.$


\section{Numerical hydrodynamical simulations}
\label{secSetup}

In this Section, we describe the set-up for our numerical hydrodynamical simulations of two-dimensional gas discs with embedded planets, using a non-barotropic equation of state. The initial  disc models and system of units are the same as those employed for the linear calculations. The softening parameter is  $b =0.03$. This approximately corresponds to the putative disc semi-thickness and  
the use of such a softening parameter approximates the result of appropriate vertical averaging of the
gravitational potential. There is no explicit viscosity in the model unless otherwise stated. In Sect. \ref{secHeatDiff} we include explicit heat diffusion. For locally isothermal simulations, we drop the energy equation and instead use a fixed temperature profile that gives rise to a constant aspect ratio in the initial state. For models including the energy equation we have used different temperature profiles to vary the entropy gradient while keeping the density gradient constant. Unless stated otherwise, we slowly build up the mass of the planet during the first three orbits to avoid to avoid transients due to the sudden introduction of the planet into the disc. For low-mass planets in isothermal simulations this is usually unnecessary, but the non-linearities in the flow associated with an entropy gradient can introduce some artifacts (see below).
 
\subsection{Code description}
We use the RODEO method \citep{2006A&A...450.1203P} to evolve the two-dimensional Euler equations in cylindrical coordinates $(r,\varphi)$. The grid consisted of 512 radial cells, equally spaced between $r=0.4$ and $r=1.8$, and 2048 azimuthal cells, equally spaced over $2\pi$. We used non-reflective boundary conditions \citep[for details see][]{2006A&A...450.1203P}. The energy equation is a straightforward extension of the method \citep[see][]{1995A&AS..110..587E}, which was also used in \cite{2006A&A...459L..17P} to perform three-dimensional, radiation-hydrodynamical simulations of embedded planets. In this paper, we have not used the radiation diffusion solver. Heat diffusion is incorporated the same way as Navier-Stokes viscosity in \cite{2006A&A...450.1203P}.

In calculating the torque on the planet, we include all disc material, including 
that within the Hill sphere of the planet. For low-mass planets considered here, this should not influence the results.

\begin{figure}
\centering
\resizebox{\hsize}{!}{\includegraphics[]{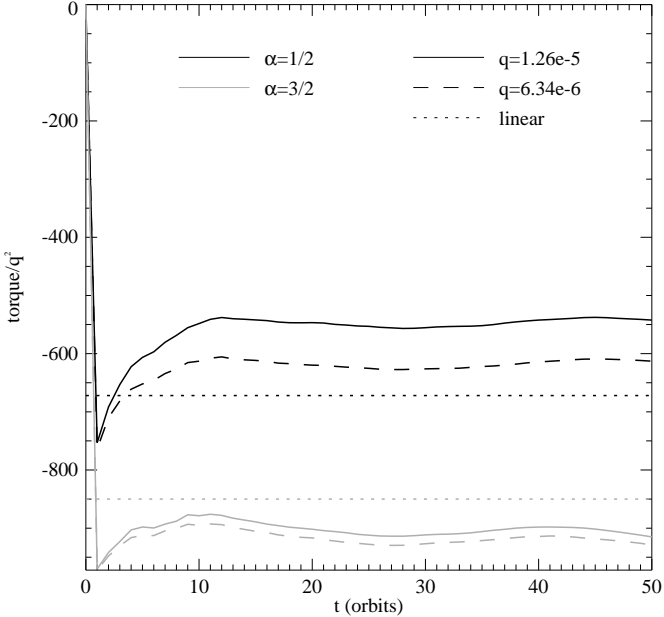}}
\caption{ Total torque in units of $q^2\Sigma(\rp)\okep^2\rp^4$ as a function of time for two different planet masses and two density profiles for the disc. The disc was kept locally isothermal, and a viscosity $\nu=10^{-5}$ $\rp^2 \op$. The dotted lines indicate the results of our linear calculations.}
\label{figisovisc}
\end{figure}

\subsection{Comparison to previous results}

Before embarking on a study of non isentropic discs in Sect. \ref{secNum},
we compare our results for locally isothermal discs with previous work and expectation from linear theory in this section.

Different kinds of comparisons of torque calculations with linear theory
have been reported in the literature. \cite{2006ApJ...652..730M}
compare locally isothermal three dimensional calculations with 
the linear results of \cite{2002ApJ...565.1257T}. This comparison is also
reported in \cite{2007prpl.conf..655P}. In this comparison
the expression for the linear torque given by
\be T_L = - (1.35+ 0.55\alpha)q^2(H/r)^{-2}\Sigma(\rp)\okep^2\rp^4\label{compare}\ee is used.
On the other hand the results of two dimensional simulations
using a softened planet gravitational potential, as is done here,
have also been compared to those given by Eq. (\ref{compare})
and it has been found that an approximate match 
may be achieved with an appropriate choice of softening parameter
\citep{2004MNRAS.350...849N}. We shall show that our results
are fully consistent with those reported in  \citet{2004MNRAS.350...849N}
 and  Eq. (\ref{compare})
 as well as our own customized
linear calculations below.
We note that the comparisons referred to above are done
with constant kinematic viscosity coefficients, $\nu = 10^{-5}r_p^2\Omega_p,$
and so we adopt this value here to make our comparisons.
We comment that for this value of $\nu$,
corotation torques are expected to be largely unsaturated.

 We have performed locally isothermal simulations of planets with different mass ratio $q$ 
for different values of $\alpha,$ $H/r=0.05$ 
 and the stated value of $\nu.$

In Fig. \ref{figisovisc} we show the total torque, divided by the mass ratio squared, as a function of time for planets with $q=6.34\times10^{-6}$ and $q=1.26\times10^{-5}$ for two different values of $\alpha.$ 
In these cases the torques attain almost steady values after about
ten orbits with any remaining transients due to the initial conditions
producing only small effects.
If the response were fully linear, there would be no differences between planets
 of different masses. It is clear that the disc with $\alpha=3/2$,
 which has no vortensity gradient, is consistent with the interaction being linear.
The torques agree to within 10 $\%$ with the value obtained from our linear calculations, as well as with that obtained from Eq. (\ref{compare}). 
For the case with $\alpha=1/2$, the differences between the two planet masses are more pronounced
We find the 
magnitude of the torque divided by $q^2$  to be about $10\%$ larger for the smaller mass
indicating the presence of weak non linear effects. For that mass the torques again  agree to within 10 $\%$ with the value obtained from our linear calculations, as well as that found using  Eq. (\ref{compare}). 

In agreement with \citet{2004MNRAS.350...849N}, torques can be fitted
by Eq. (\ref{compare}) to within $10-15\%$ for the
mass range considered for an appropriate softening
parameter. In our case this would be close to $b=0.6H/r$ the value we used for almost all cases. \citet{2004MNRAS.350...849N} adopted $b=0.7H/r,$ which is not a significant difference in this context.

\section{The  development of positive corotation torques resulting from
initial entropy gradients} 
\label{secNum}
In this section we focus on the development of positive torques arising from
initial entropy gradients within the first horseshoe libration period, dealing with 
the possibility of sustaining such torques over longer time scales in 
later sections. We will consider various disc models, ranging from one with $\alpha=-1/2$, which has strong gradients in entropy and vortensity and a surface density that increases outwards, to discs with constant entropy and vortensity and outwardly decreasing surface density.

\subsection{An illustrative case}
\begin{figure*}
\includegraphics[width=\textwidth]{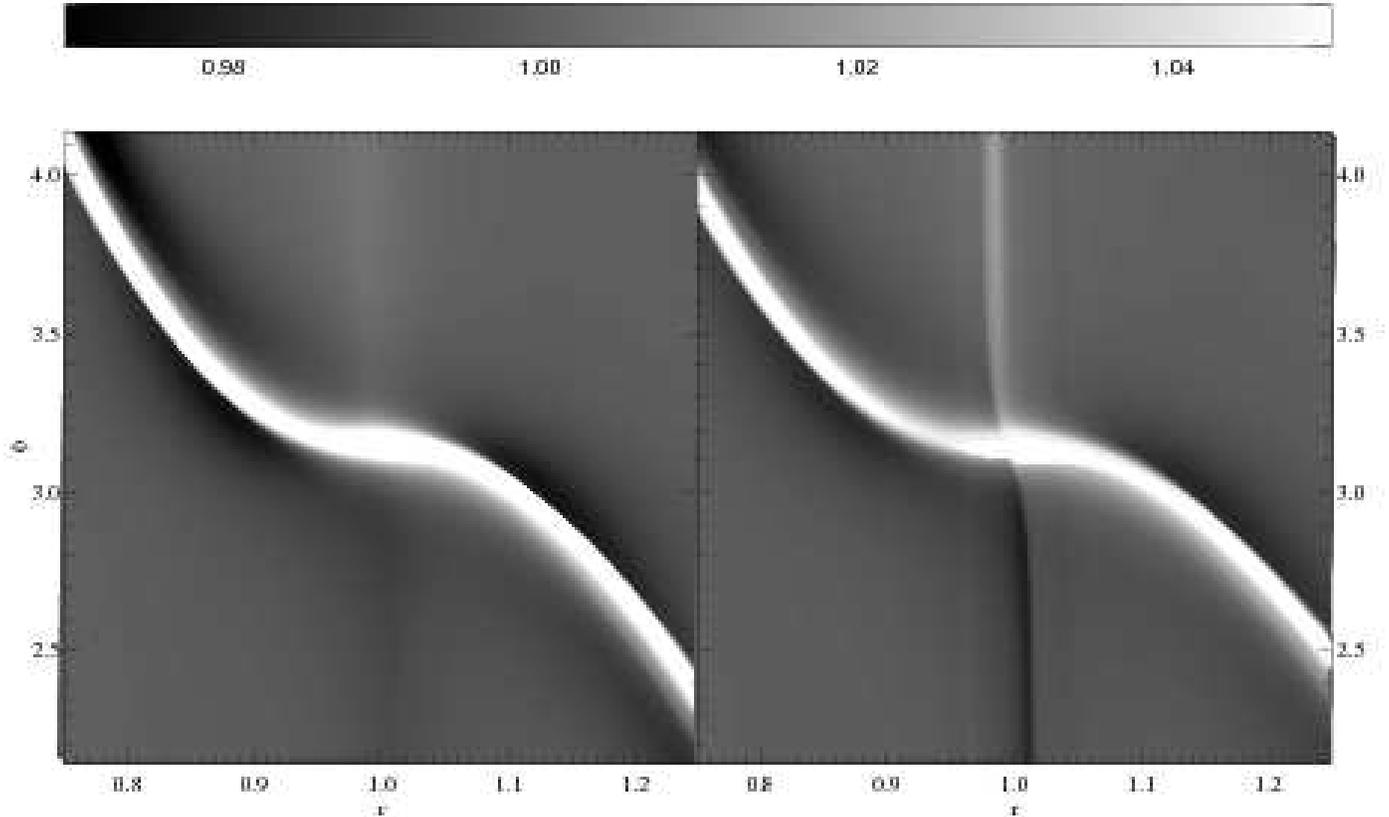}
\caption{Overview of $\rho r^{1/2}$ for a $q=1.26\cdot 10^{-5}$ planet (4 \mearths around a Solar mass star) after 20 orbits. Left panel: locally isothermal equation of state. Right panel: adiabatic equation of state.}
\label{fig1}
\end{figure*}

We  begin  by considering a planet with  $q=1.26\cdot 10^{-5}$  (corresponding to a $4$ $\me$ planet  around a Solar mass star). This  closely resembles the case studied in \cite{2006A&A...459L..17P}. 
The initial density and temperature structure is characterized by $\alpha=1/2$  and constant $H/r.$ The power law index for the entropy is $\xi= d(\log P/\rho^{\gamma})/d\log r=-0.8,$
which corresponds to $\gamma = 1.4.$ In Fig. \ref{fig1} we give contour plots of  the density (multiplied by $r^{1/2}$) for an isothermal equation of state and an adiabatic simulation after 20 orbits. The adiabatic case can be compared with the plot in Fig. \ref{fig0} obtained from the linear theory with some heat diffusion added to regularize the density perturbation. 

The wave structure differs slightly between the  simulated adiabatic and isothermal cases. Note that the color scale has been chosen to highlight corotation features, but still the differences between the spiral waves can be seen. The pitch angle differs such that the waves are more open and weaker in the adiabatic case. This is due to the fact that linear waves are isentropic, leading to a higher sound speed and less compression for equal forcing in the adiabatic case. This results in a reduction in the strength of the Lindblad or wave torques 
in adiabatic simulations compared to isothermal ones. 

The two simulations in Fig. \ref{fig1} also differ around corotation (near $r=\rp$). The adiabatic simulation shows a strong, narrow density feature
which is associated with a positive perturbation (light shade) for $\varphi > \pp$  
and associated with  a negative perturbation (dark shade) for $\varphi < \pp.$
 As the planets are moving upwards in Fig. \ref{fig1}, it is clear that this feature acts as to accelerate the planet in its orbit, causing \emph{outward} migration. It is also found in the linear calculations but at reduced strength (see Fig. \ref{fig0}). In Sect. \ref{secNonlin} we will see that this is because linear theory fails to take into account the horseshoe turns. We will see below that when non-linear simulations are undertaken, this corotation torque can easily come to dominate the Lindblad torques, and that the sign depends on the direction of the entropy gradient in the disc. The isothermal simulation also shows a corotation feature, which is due to the radial vortensity gradient in the disc. Although the sign of this corotation torque is positive, it is not strong enough to overcome the negative wave torque. For planets of higher mass the situation   may be different \citep{2006ApJ...652..730M}.   
  
\begin{figure}
\centering
\resizebox{\hsize}{!}{\includegraphics[]{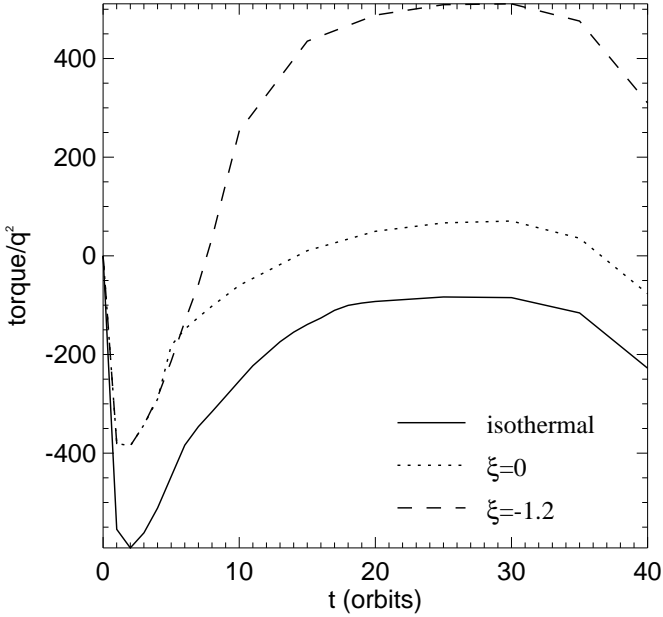}}
\caption{Total torque in units of $q^2\Sigma(\rp)\okep^2\rp^4$ as a function of time for a $q=1.26\cdot 10^{-5}$ planet (4 \mearths around a Solar mass star), using a locally isothermal equation of state (solid line) and solving the full energy equation, for two different radial entropy gradients 
$\xi=d\log(P/\rho^{\gamma})/d\log r$ (dotted line: $\xi=0$, dashed line: $\xi=-1.2$). All models have $\alpha=-1/2$.}
\label{fig2}
\end{figure}

\begin{figure}
\centering
\resizebox{\hsize}{!}{\includegraphics[]{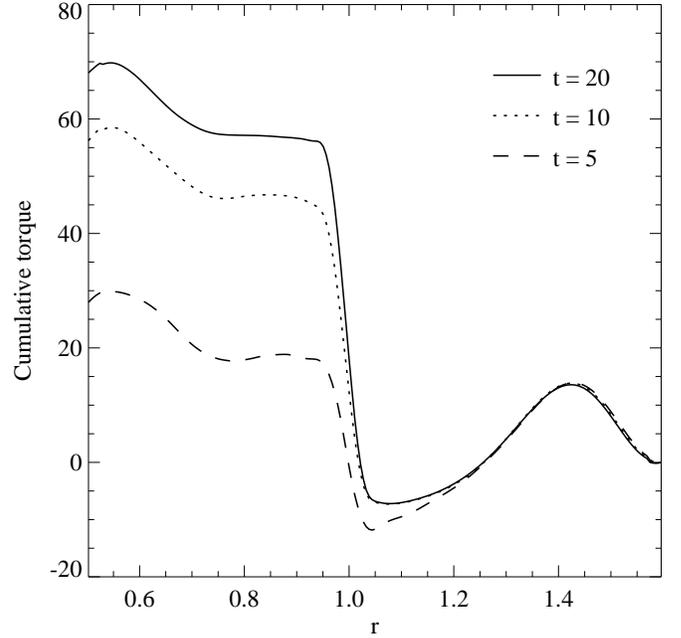}}
\caption{The  cumulative torque in units of $ q^2 \Sigma(\rp)\okep^2\rp^4$ acting on the protoplanet plotted as a function of dimensionless radius for $\rho_0 \propto r^{1/2},$ $\xi = -1.2$  
and $m=2$ at different times.  The cumulative torque is defined to be zero at the outer boundary and equal the total torque at the inner boundary. 
}
\label{fig3}
\end{figure}

In Fig. \ref{fig2} we compare the time evolution of the total torque for an isothermal and an adiabatic simulation with $\alpha=-1/2$, together with a run with the same density profile, but with a temperature profile such that the entropy is constant.  In the case with a negative entropy gradient, we find a positive torque on the planet after approximately 10 orbits. After that time, the torque continues to rise steadily in all cases. This is when the corotation torque is set up, see Fig. \ref{fig3}
which shows the torque associated with the $m=2$ component of the
density perturbation for the case with negative entropy gradient. The torque difference between 20 and 10 orbits arises entirely at corotation. In all cases shown in Figs. \ref{fig2} and \ref{fig3}, the corotation torque is expected to be positive due to the positive vortensity gradient and negative (or zero) entropy gradient.

However, in linear theory the magnitude of the corotation torque has been found to be  not enough to produce a positive torque and thus smaller than the wave torque. We note that  one finds from 
\cite{2002ApJ...565.1257T} that the positive linear corotation torque for $\alpha = -1/2$ in the isothermal case is expected to be  $54\%$ of the wave torque in magnitude making the total torque $46\%$ of the latter. Also, the linear torques are expected to reach their final value on a dynamical time scale, {\it  which is independent of the planet mass ratio}, rather than over $20$ orbits as is seen in Fig. \ref{fig2}. The precise time to reach the maximum depends on the time scale on which the planet is introduced, but it is always   significantly more than a dynamical time scale. Below, we will see that this time depends on the planet mass ratio and so  this behavior must be associated with non-linearities in the flow  and we interpret it as arising from torques generated after  horseshoe turns as described in Sect. \ref{secNonlin} below.

After approximately $35$ orbits, the torque starts to drop again in all cases. This is also an indication that corotation torques play a major role, because it is expected that they start to saturate after half a libration period, which is close to $35$ orbits for this planet.  We note that the torque then
 starts to oscillate on the libration time scale  and, ultimately after the action of  non-linear effects and diffusion, the disc settles down to a state with zero entropy and vortensity gradients around corotation in an average sense. Therefore, the corotation torque vanishes unless some form of  diffusive or evolutionary process can restore the entropy gradient within one libration period. We will come back to this issue in Sect. \ref{secLongterm}.

\begin{figure}
\centering
\resizebox{\hsize}{!}{\includegraphics[]{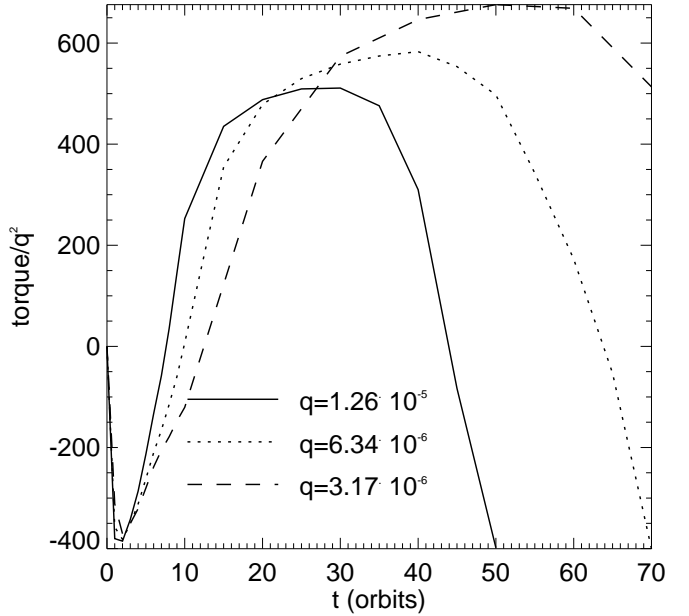}}
\caption{Total torque in units of $q^2 \Sigma(\rp)\okep^2\rp^4$ as a function of time for 3 different planet masses and an adiabatic equation of state with $\xi=-1.2$ and $\alpha=-1/2$. The torque is divided by $q^2$ to filter out purely linear effects.}
\label{fig4}
\end{figure}

\begin{figure}
\centering
\resizebox{\hsize}{!}{\includegraphics[]{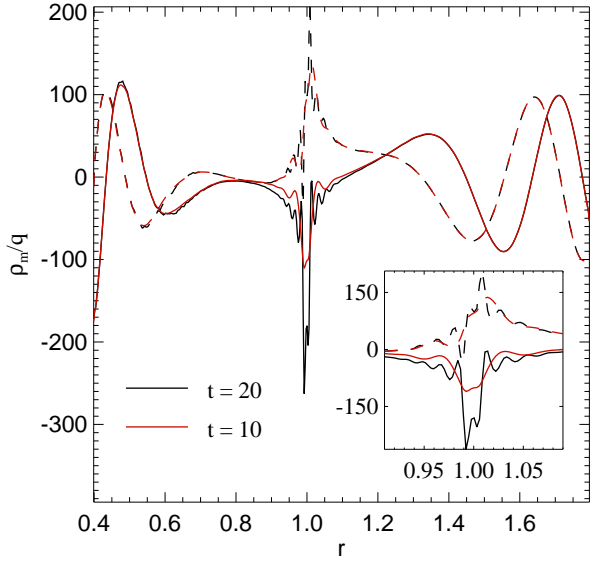}}
\caption{Real (dashed lines) and imaginary (solid lines) parts of the $m=2$
Fourier component of the density perturbation due to a
$q=3.17\cdot 10^{-6}$ planet after 20 (black) and 10 (red) orbits in an adiabatic disc with $\xi=-1.2$ and $\alpha=-1/2$. The inset shows a close-up on the corotation region. A comparison  with results obtained from linear theory indicates the form of a linear response apart from in the corotation region.}
\label{fig7}
\end{figure}

 \begin{figure}
 \centering
\resizebox{\hsize}{!}{\includegraphics[]{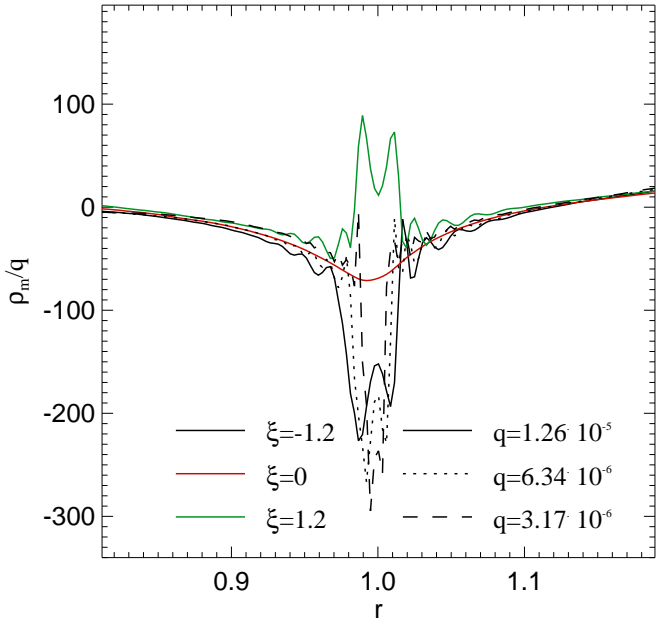}}
 \caption{Imaginary parts of the $m=2$ Fourier component of the density perturbation due to a $q=1.25\cdot 10^{-5}$ planet for different radial entropy profiles with $\alpha=-1/2$. For the negative entropy slope, we show three different planet masses.}
\label{fig8}
\end{figure}

\begin{figure}
\centering
\resizebox{\hsize}{!}{\includegraphics[]{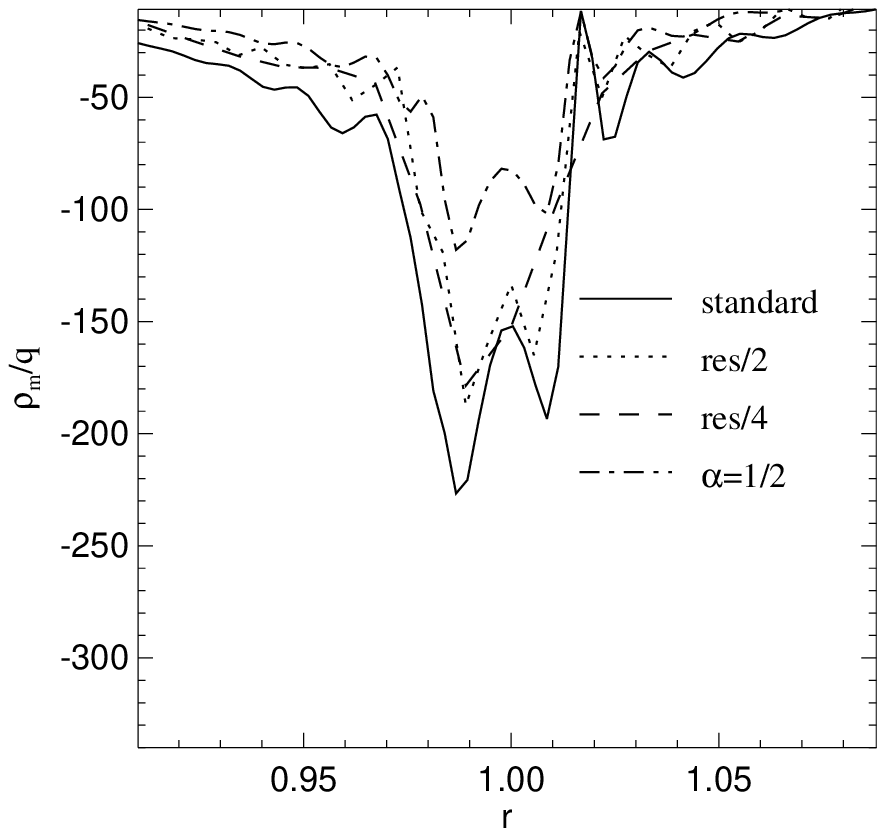}}
\caption{Imaginary parts of the $m=2$ Fourier component 
of the density perturbation due to a $q=1.25\cdot 10^{-5}$ 
planet for different resolutions with $\xi=-1.2$ and $\alpha=-1/2$, together with a run with a steeper radial density profile ($\alpha=1/2$, with the same value for $\gamma$).}
\label{fig9}
\end{figure}

\subsection{Non-linear evolution}
Even at early times the evolution of the torque is dominated by non-linear effects. In Fig. \ref{fig4} we show the time evolution of the torque for three different planet masses.  Torques have been scaled by $q^{-2}$, so that if the response was  purely linear all curves should lie  on top of each other.  The time scale for the rise  towards positive values clearly depends on the  mass ratio $q,$ which would not be the  case if the effect was linear. The time to reach the same degree of non-linearity is consistent with being proportional with $q^{-1/2}$, as implied by Eq. \eqref{cond2}.
Similarly, the time scale on which saturation sets in, resulting in the torque dropping  below zero again, depends on $q.$  Therefore, although there is a linear effect associated with the entropy gradient,  the phenomenon that results in the torque switching  sign is clearly non-linear. 

This is further illustrated in Fig. \ref{fig7}, where we show the $m=2$ component 
of the surface density perturbation due to a $q=3.17\cdot 10^{-6}$ planet 
(corresponding to a $1$ $\me$ planet  around a Solar mass star) 
after $10$ and $20$ orbits. The imaginary parts (solid lines) again are directly 
related to the torque. At the boundaries of the domain we see clear signatures of outgoing waves. 
The difference between $t=20$ and $t=10$ arises near corotation only (see also Fig. \ref{fig3}). At $t=10$, the perturbation is reasonably smooth and  resembles the fully developed linear case away from corotation 
(see Fig. \ref{fig01}).  
As time progresses, however, the corotation feature becomes narrower and deeper, 
until it reaches a final width. 
This finite final width is again indicative of non-linear behavior, and from Fig. \ref{fig7} 
it is clear that this non-linear structure is responsible for the cumulative torque on the planet  reaching positive values. The ripples on the edges of the corotation region in Fig. \ref{fig7} are of numerical origin and have to do 
with the strong non-linearity in the density gradient. They can be reduced by slowly ramping up the mass of the planet (which was done for the runs in Fig. \ref{fig7}) but in the absence of heat diffusion they never go away completely.  

In Fig. \ref{fig8} we show the imaginary part of the $m=2$ Fourier component of the density for the three planet masses (black lines). These allow us to make an easy estimate of the width of the horseshoe region $\xs$. Interestingly, if we define the horseshoe region to extend to the distance where the torque is half the minimum, we find that $\xs=\rp\sqrt{2q/3b}$ to within a few percent. 
Also, it can be deduced from  Fig. \ref{fig4} that the time scales for which saturation 
sets are consistent with libration period $8\pi/(3\xs)$  set by  this value, which one may obtain
from considering the gravitational and centrifugal equipotentials alone. In this context we comment that
from a consideration of horseshoe streamlines, \citet{2006ApJ...652..730M} estimated   $\xs \sim \rp\sqrt{8\rp|\phigps|/(3GM_*)},$ where $\phigps$ is the gravitational potential of the planet evaluated at the point  on the horseshoe separatrix closest to the planet (as we use this expression only  to  indicate a
scaling, we neglect an additional enthalpy contribution to the potential). According to this we expect 
 $\xs=\rp\sqrt{2q\rp/(3L)}$ for an appropriate length scale $L.$  As \citet{2006ApJ...652..730M} point out, in practical cases,  this should be determined by both the softening length and the scale height, $H.$
 But note that the softening length is generally chosen so that these are closely related.
For example these authors  provide   the estimate  
 \begin{equation} 
\xs=1.16 \rp\sqrt{q\rp/H}.\label{Masxs}
\end {equation}
 This corresponds to $L=0.495H$
 which therefore  agrees  with the  estimate above which used $L=0.6H=b$  to within  ten percent. 
They indicate that
their estimate fits locally isothermal simulation results at smaller softening lengths and that this should be used
to give $x_s$  when $b\rightarrow 0.$

Incorporating the above ideas, following \citet{2006ApJ...652..730M} from now on   we define the horseshoe width through 
\begin{equation}
\xs=\lambda \rp \sqrt{\frac{2q \rp |\phigps|}{3G\mpl}},
\label{eqxs}
\end{equation}
where $\lambda$ is expected to be of order unity and can be determined from simulations.

\begin{figure}
\centering
\resizebox{\hsize}{!}{\includegraphics[]{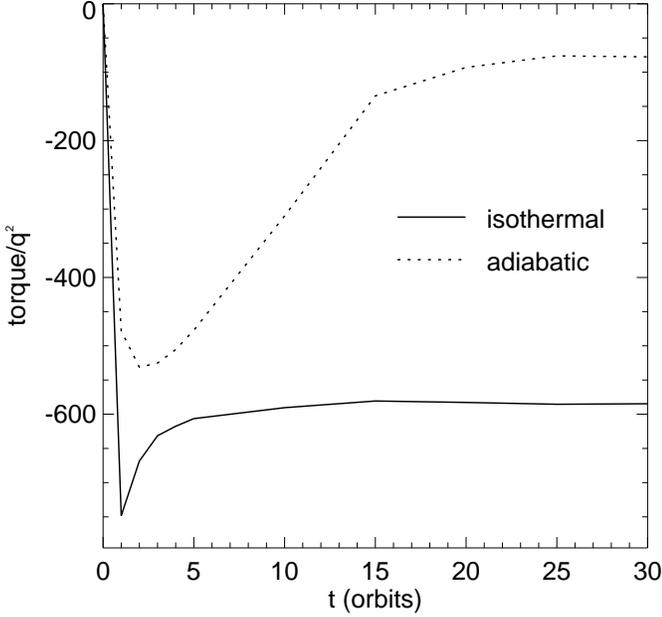}}
\caption{Total torque in units of $q^2 \Sigma(\rp)\okep^2\rp^4$ as a function of time for $\rho_0 \propto r^{-1/2}$ and $q=1.26\cdot 10^{-5}$. The solid line shows an isothermal run, the dotted line an adiabatic run.}
\label{fig10}
\end{figure}

The total torque, divided by $q^2$, depends on the mass of the planet (see Fig. \ref{fig4}). The reason for this is at least partly through the form of the density profile around corotation. For the torque to be proportional to $q^2,$ we would expect that the mass associated with the feature, or any of its Fourier components, should scale as $q.$ From Fig. \ref{fig8}, both  the width and the depth of the feature are seen to depend on mass but in such a way that the area increases more weakly than $q^2.$ But note that the small scale structure apparent in the density perturbation around corotation may well depend strongly on numerics, and therefore also the total torque. We will show below that this problem  can be overcome by adding a small amount of thermal diffusion.

In Fig. \ref{fig9} we show the corotation feature for three different resolutions. It appears that the corotation feature gets stronger with increasing resolution, and it is not clear if the torque has converged for our highest resolution. A common feature for all runs is the positive total torque. The total torque is also remarkably robust to variations in the flux limiter, showing differences of less that 5\% between the minmod limiter and the soft superbee limiter \citep[see][for their definition]{2006A&A...450.1203P}. From Fig. \ref{fig9} it is clear that there are strong features in the density that have a width of one grid cell for all resolutions. At the two highest resolutions, we resolve all relevant scales: the pressure scale height, the Hill radius of the planet and the width of the horseshoe region. The appearance of these small scale features even at these high resolutions suggests that the torque is limited by numerical diffusion. We will   return to this issue below, where we discuss models with explicit heat diffusion. 

\cite{2006ApJ...652..730M} noted non-linear behavior in the barotropic case above a certain mass. Our analysis suggests that in the non-barotropic case, for sufficiently low thermal diffusivity, and for sufficiently long integration times, non-linear effects are always dominant (see Eqs. \eqref{cond1} and \eqref{cond2}). In addition, the discussion of the linear problem indicates that, as  the state variables (but not their gradients) converge as $\epsilon \rightarrow 0,$ non-linear effects should be much milder in the barotropic case. This is confirmed by Fig. \ref{fig2}, where  the non-linear evolution takes the torque on the planet to positive values only in the non barotropic case. 

In Fig. \ref{fig8}, we  show the results for different entropy gradients, with the planet mass ratio  being fixed at $q=1.26\cdot 10^{-5}$. The run with no entropy gradient resembles the Lorentz profile of the linear case (see the right panel of Fig. \ref{fig01}), confirming that it is the entropy gradient that drives the non-linearity. A positive entropy gradient changes the sign of the corotation feature, making the corotation torque on the planet negative. The run with no entropy gradient is intermediate between the positive and negative entropy gradient runs and indicates a linear dependence of this torque on $\xi$. However, because we measure total 
torques arising from a combination of physical processes, all of which vary
when the disc structure is modified by altering the entropy gradient,
this is not conclusive, because entropy-related torque contributions may not always be reliably identified.

The run for which we show the corotation feature in Fig. \ref{fig9} behaves differently when $\alpha$ 
is changed from $-1/2$ considered so far to $\alpha=1/2$. 
In Fig. \ref{fig10} we show the evolution of the total torque, comparing an isothermal 
run with an adiabatic run with $\xi=-0.8$. For this disc, the total torque never becomes positive,a
 indicating a strong dependence of the migration rate on the density profile of the unperturbed disc. 
In fact, there are three effects
 conspiring to make the torque negative in the $\alpha=1/2$ case. First of all, the wave
or Lindblad torques  are expected to be 
  almost the same being only slightly weaker  \citep[see][]{2002ApJ...565.1257T}. Second, the 
positive contribution from the  linear barotropic corotation torque, that would apply to the locally isothermal case 
if the temperature variation across the corotation region may be neglected, is expected to be weaker by 
a factor of two  because the radial vortensity gradient is less strong\footnote {Note that in a cylindrical disc the vortensity gradient is zero for $\rho_0 \propto r^{-3/2}$ (see Eq. \eqref{eqdefvort}).} \citep{1979ApJ...233..857G}. 
Not only will the linear torque be more negative, there is also no expected strong non-linear boost for the barotropic corotation torque in these circumstances. This is clear from the solid line in Fig. \ref{fig10}, which does not show the strong rise that was apparent in Figs. \ref{fig2} and \ref{fig4}. For this example the final torque is also in reasonable agreement with that obtained from the linear analysis which is also indicative of less important non-linear effects at corotation. Finally the radial entropy gradient is less negative than in the case with $\alpha=-1/2$, which reduces the positive contribution of the non-barotropic corotation torque. The result is a negative torque on the planet, and therefore inward migration, albeit at a slower rate due to the positive contribution of the entropy gradient to the torque. 
 
\begin{figure}
\centering
\resizebox{\hsize}{!}{\includegraphics[]{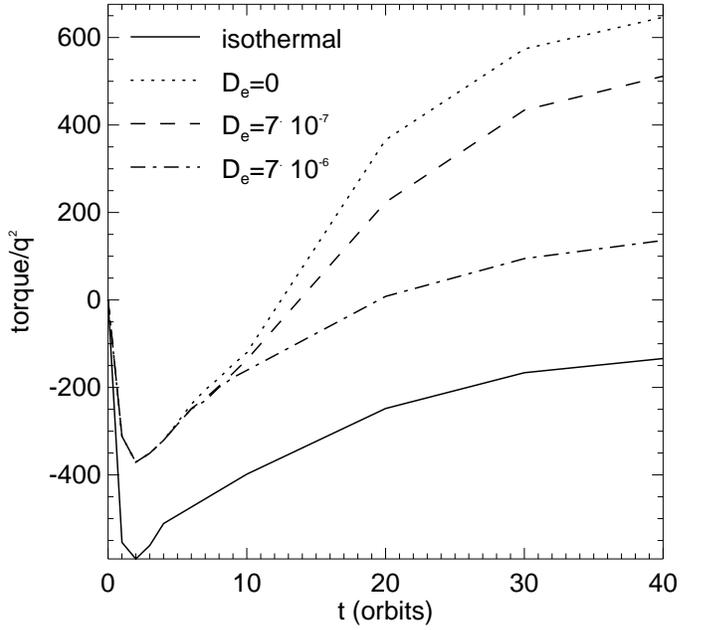}}
\caption{Total torque in units of $q^2 \Sigma(\rp)\okep^2\rp^4$ as a function of time on a $q=3.17\cdot 10^{-6}$ planet for different values of the heat diffusion coefficient $\de$, in a disc with $\alpha=-1/2$ and $\xi=-1.2$.}
\label{fig11}
\end{figure}

\begin{figure}
\centering
\resizebox{\hsize}{!}{\includegraphics[]{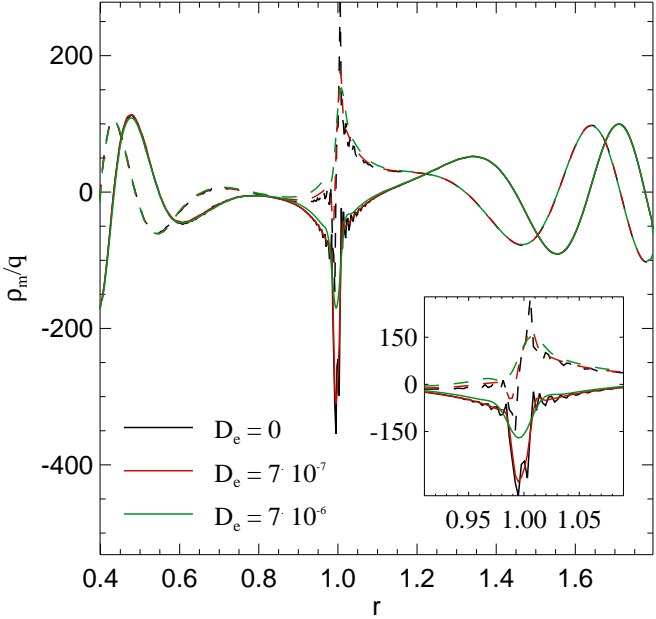}}
\caption{Real (dashed lines) and imaginary (solid lines) parts of the $m=2$ Fourier component of the density perturbation due to a $q=3.17\cdot 10^{-6}$ planet after 40 orbits, for different values of the heat diffusion coefficient $\de$, in a disc with $\alpha=-1/2$ and $\xi=-1.2$. The inset shows a close-up on the corotation region.}
\label{fig12}
\end{figure}

\begin{figure}
\centering
\resizebox{\hsize}{!}{\includegraphics[]{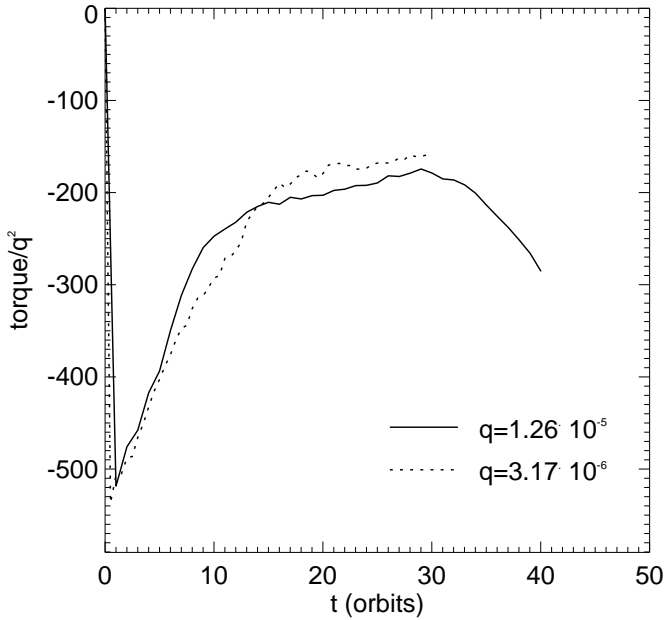}}
\caption{Total torque in units of $q^2 \Sigma(\rp)\okep^2\rp^4$ as a function of time for $\rho_0 \propto r^{-1/2}$
 and two different planet masses. For the $q=3.17\cdot 10^{-6}$ planet,
 we have used $\de=7\cdot 10^{-7}$, for the $q=1.26\cdot 10^{-5}$ 
planet we have used $\de=5.6\cdot 10^{-6}$.}
\label{fig13}
\end{figure}

\subsection{Effect of heat diffusion}
\label{secHeatDiff}
In this Section we add a small thermal diffusivity into the model, characterized by the value of $\de$. We expect that for a  high enough value of $\de,$ in the vicinity of the corotation radius, all non-linear behavior due a radial entropy gradient disappears and the model should behave as if it was locally isothermal. In Fig. \ref{fig11} we show the time dependence of the total torque on a $q=3.17\cdot 10^{-6}$ planet for different values of $\de$ with $\xi=-1.2$ and $\alpha=-1/2$. It is clear that heat diffusion decreases the strength of the non-linear part of the torque. However, for times less than $5$ orbits, the behavior is independent of $\de.$ This is consistent with the solution being in the linear regime during this phase (see Sect. \ref{secLin} above).

For $\de=7\cdot 10^{-7}$ the torque during the later non-linear phase is still positive, but less so than in the adiabatic case ($\de=0$). Thus this value of $\de$ is enough to affect the non-linear solution, which is consistent with the estimates made in Sect. \ref{secLin} (see Eq. (\ref{cond1})). Increasing $\de$ by a factor of $10$ gives an even  weaker positive torque on the planet. The fact that after $5$ orbits there is a constant shift in torque between this case and the isothermal case indicates that all non-linear effects due to the entropy gradient have disappeared. The difference between these cases is simply due to  different Lindblad or wave torques, which is supported by the fact that after $5$ orbits the torque ratio is about equal to $\gamma.$ Together with the mildly non-linear barotropic part of the corotation torque 
(which is not affected by heat diffusion, but which may still involve significant density gradients in the coorbital region, see Sect. \ref{secLongterm}) this is enough to switch the sign of the torque to positive values. 

To illustrate the behavior of the entropy-related torque, we show the $m=2$ Fourier component of the density perturbation in Fig. \ref{fig12} for the cases $\de=0$, $\de=7\cdot 10^{-7}$ and $\de=7\cdot 10^{-6}$. It is clear that the effects of heat diffusion are localized around corotation, where the temperature gradients are strongest. The close-up around corotation shows a smooth Lorentz-profile for the most diffusive run, resembling the linear case in a qualitative way (see the right panel of Fig. \ref{fig01}). Note that Eq. \eqref{cond1} predicts that non-linear effects should be present for $\de<3.4\cdot 10^{-7}$, while from  Fig. \ref{fig11} we see that for a diffusion coefficient that is twice as high non-linear effects associated with the entropy gradient are still present. Bearing in mind that Eq. \eqref{cond1} gives only a rough estimate, and that non-linear effects may indeed set in for a somewhat  lower values of $\de,$ the agreement is reasonable. 

Even in the less diffusive case the sharp features at corotation that can be seen in Fig. \ref{fig12} for the
 $\de=0$ case are smoothed out. In fact, for $\de=7\cdot 10^{-7}$ we find the same density response for a simulation with half the resolution, and also the total torques agree to within 5 $\%$.  Therefore, the torque is numerically well-defined, unlike the case with $\de=0.$ If we now compare different masses with a minimum amount of diffusion, enough to smooth out the sharp features but at the same time keep the non-linear behavior, we find that the total torque scales almost as $q^2$. Note that in order to have the same amount of non-linearity in the problem, we need to use different values of $\de$ for different masses (see Eq. \eqref{cond1}). The results are shown in Fig. \ref{fig13} for $\alpha=1/2$ and $\xi=-0.8$, where we have rescaled the time for the low-mass planet by a factor of 2. This removes the $q^{-1/2}$ scaling for the time to set up the torque. Interestingly, the torque plateau in these cases is reduced by a factor of $2$ compared to the case with $\de =0$, being negative and about $1/2$ of the linear value. This analysis supports the idea that the departure from the scaling of the torque with $q^2$ is due to non-linear effects in the corotation region.


\section{A simple non-linear model}
\label{secNonlin}
Because linear theory always breaks down for sufficiently small $\de$, it is necessary to develop a non-linear picture to understand the results of the numerical simulations presented above in Sect. \ref{secNum}. In this section, we discuss a very simple non-linear model that can give at least  qualitative insight into the torque behavior. It is a non-linear  model because  a finite width of the corotation region, in which the gas travels on horseshoe orbits, is adopted.
Such a region is {\it not} present in linear theory.
 The model only accounts for the breakdown of linear theory and the build-up of a non-linear
 contribution to the corotation torque in the early stages  after a planet is inserted into the disc.
 The later expected saturation of the torque is not considered.

\begin{figure}
\centering
\resizebox{\hsize}{!}{\includegraphics[]{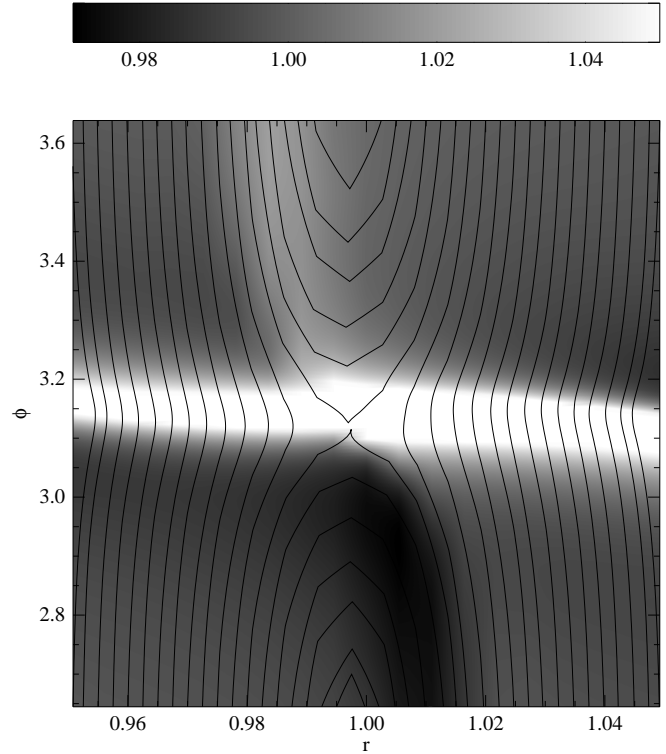}}
\caption{Density, multiplied by  $r^{3/2}$, after 10 orbits of a $q=1.26\cdot 10^{-5}$ planet embedded in a disc with $\rho_0 \propto r^{-3/2}$, $T \propto r^{-1}$ and $\xi=-0.85$. Approximate locations of the streamlines are indicated by the black lines.}
\label{figstreamline}
\end{figure}

A description of the way the model works is as follows. Viewed in the frame rotating with the planet, material exterior to the planet approaches  the azimuth of the planet along approximately linear trajectories. It executes a horseshoe turn, subsequently moving away from the planet when it will eventually
will have received a radial displacement equal to twice the initial distance from the orbital radius of the planet (see the streamlines plotted in Fig. \ref{figstreamline}). If the planet were absent material would shear past without a turn. Thus the effect of the horseshoe turns is to  produce physical changes in the material that correspond to a radial displacement equal to the width of the horseshoe region after it has encountered the planet. The horseshoe region is separated from the remainder of the disc by a separatrix.

When the material conserves entropy as it moves, such a displacement produces a density change. For example, if the disc entropy per unit mass decreases outwards, when material is displaced inwards, the density will increase. Here we shall assume the displacement occurs such that fluid near separatrices
in the horseshoe region maintains pressure balance with its surroundings 
and by extension that the Eulerian change in the pressure may be neglected.
 In this context we comment that it is found both from linear theory and non-linear simulations that the relative density changes always greatly exceed the relative pressure changes (see above and below).
 We note that the dynamics we have described are clearly seen in the simulations. In Fig. \ref{figstreamline} we show a close up of the corotation region near the planet for a 4 $\me$ planet.
The density structure indicates an increase inside the horseshoe region that is bounded by the inner separatrix. This is also indicated by the form of the streamlines superposed on the plot. Note that these are slightly distorted by the grid resolution.
  We comment that the change in density that occurs after a horseshoe turn is expected to result in a slight asymmetry in the positions of the horseshoe legs on either side of the corotation radius away from the planet. But for the relative  density changes considered here (typically of order one percent) this is a very minor effect. 

The horseshoe turns commence immediately after a planet is inserted into a disc and produce a torque that builds up over time in addition to that expected from linear perturbation theory.  Departures from  linear theory are thus produced early on. As regions close to the planet are the most important in establishing the torque, a plateau is reached fairly quickly when these regions attain a quasi-steady state. This lasts   for the order of a libration period after which material recirculates to the location of the planet and the torque is expected to saturate. The model described here accounts for the attainment of the torque plateau but not the final saturation.

To describe the model quantitatively, we use coordinates $(x,\theta)$ that are centered on the planet: $x=(r-\rp)/\rp$, $\theta=\varphi-\pp$. As a fluid element turns to move from one
side of the planet to the other, once it has moved there is ultimately a jump from $x$ to $-x$; after this the streamlines are along lines of constant $x$. Here we neglect minor effects 
that may arise from the slight offset of the centre of the horseshoe region
from the planet's orbital radius due to the background pressure 
gradient.

For simplicity, we concentrate on the part of the corotation torque that is due to the entropy gradient only.
In general one also expects  an additional contribution if the gradient 
of specific vorticity is non-zero. 
Because most of the torque arises close to the planet we expect the major part of it
to be established before a full libration period and that this could be determined locally. Below we estimate the maximum torque arising from the initial horseshoe turns
as described above.

Because entropy is conserved along streamlines and we assume the change Eulerian in pressure is zero, we can write for the perturbed density (we focus on a jump from the horseshoe leg exterior to the planet that has $\theta>0$ to the corresponding leg interior to the planet, located in the upper segment of Fig. \ref{figstreamline}):

\begin{equation}
\rho'= -2x\frac{\xi}{\gamma}\rho_0,
\end{equation}
where $\xi=d \log(P/\rho^{\gamma})/d\log r$. The torque on the disc can be found by integrating over the disc domain that has participated in a horseshoe turn:
\begin{equation}
\mathcal T = -2H\rp^2 \int \rho' \frac{\partial \phigp}{\partial \theta}d\theta dx.
\end{equation}
The integral over $\theta$ is made simple by adopting as 
    integration limits  $ \theta = 0$ and $\theta$ large enough such that the force due to the planet
is small, which leaves us with:
\begin{eqnarray}
\mathcal T=4H\rho_0\rp \frac{\xi}{\gamma}\mathrm{G}\mpl \int_0^{\xs}\frac{1}{\sqrt{x^2+b^2}}xdx,
\end{eqnarray}
where we have approximated the planet potential by:
\begin{equation}
\phigp = -\frac{\mathrm{G}\mpl}{\rp\sqrt{x^2+\theta^2+b^2}}.
\label{plapot}
\end{equation}
Thus we  arrive at:
\begin{equation}
\mathcal T=4\mathrm{G}\mpl  H \rho_0 \frac{\xi}{\gamma}\frac{\xs^2}{\rp b+\sqrt{(\rp b)^2+\xs^2}}.
\label{eqTdisknonlin}
\end{equation}
The torque on the planet has the opposite sign and may be written in the form
\begin{equation}
\mathcal{T}_\mathrm{p}=
-\frac{4\xi \xs^2}{\gamma q \left( \rp b+\sqrt{\xs^2+\rp^2b^2}\right)} q^2\Sigma_\mathrm{p}\rp^3 \op^2,
\label{eqnonlin}
\end{equation}
where we have introduced a factor of 2 to account the additional
contribution arising from the horseshoe legs on the other side of the planet.
Thus we assume a symmetry such that 
these behave in a corresponding manner but with a density deficit
rather than increase produced near  the separatrix after a turn,
so also producing a positive torque.

We comment here that the above torque depends on the softening parameter,
$b$ through evaluation of the planet potential at the point on the separatrix closest to the planet,
here taken to be the actual planet location. 
We may  accordingly replace it with the magnitude of this potential
denoted by $|\phigps|= G\mpl/(\rp b).$ Then the torque may be written in the form
\begin{equation}
\mathcal{T}_\mathrm{p}=
\frac{-4\xi \xs^2 |\phigps|}{\gamma q G\mpl \rp
\left( 1+\sqrt{\frac{\xs^2 |\phigps|^2}{(G\mpl)^2}+1}\right)} q^2\Sigma_\mathrm{p}\rp^4 \op^2.
\label{eqnonlin1}
\end{equation}

In this form the expression for the torque may be used in the situation where
 the  point on the separatrix closest to the planet
is separated {or offset} from the planet position by an angular extent $\theta_\mathrm{s}.$
From Eq. (\ref{plapot}) we see that this is equivalent to replacing $b$
by $\sqrt{b^2+\theta_\mathrm{s}^2}.$ Now we use Eq. (\ref{eqxs}) to substitute for $\xs.$
The expression for the torque  then  becomes
\begin{equation}
\mathcal{T}_\mathrm{p}=\frac{-\frac{8}{3}\frac{\xi}{\gamma}\left(\frac{\rp|\phigps|}{G\mpl}\right)^2\lambda^2}
{1+\sqrt{\lambda^2\frac{2q}{3}\left(\frac{\rp |\phigps|}{G\mpl}\right)^3+1}}
 q^2\Sigma_\mathrm{p}\rp^4 \op^2,
\label{eqnonlin2}
\end{equation}
where $\lambda$ is the dimensionless parameter defining the horseshoe width that
 is expected to be of order unity that was introduced above 
(see Eq. (\ref{eqxs})).

\begin{figure*}
\includegraphics[width=\textwidth]{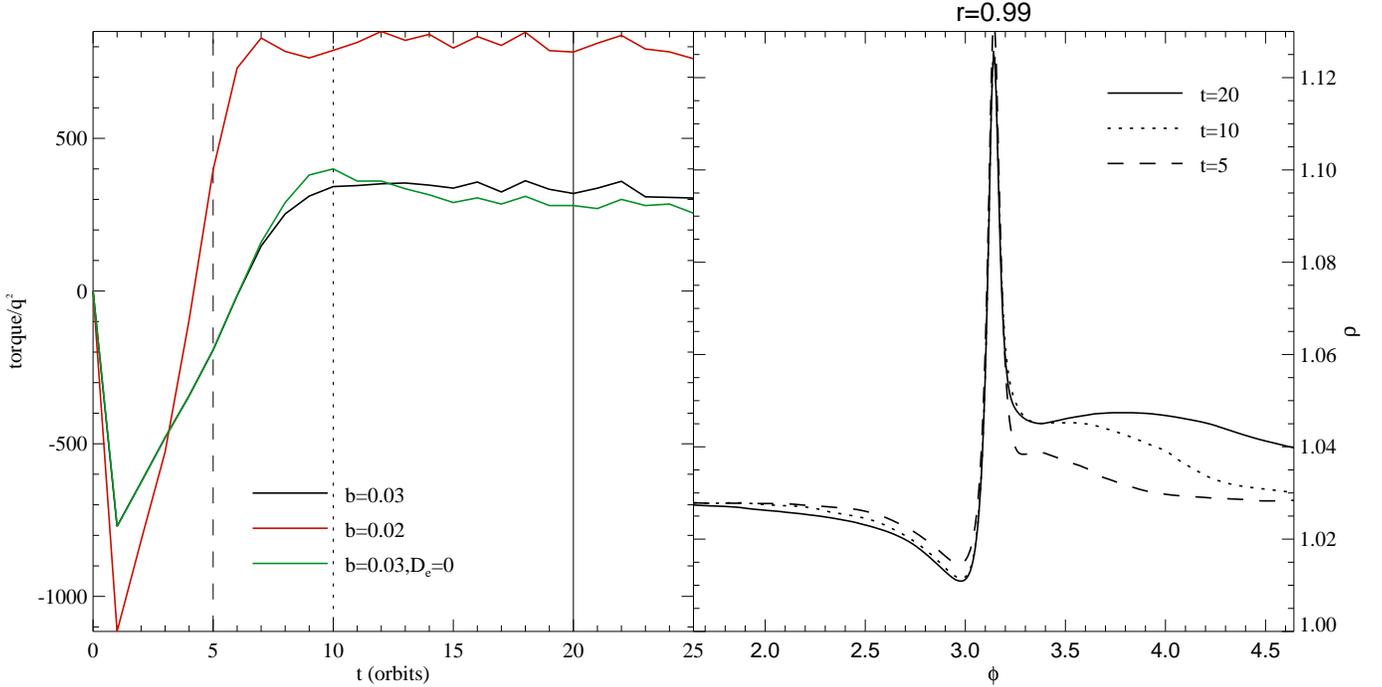}
\caption{Left panel: total torque, in units of $q^2 \Sigma(\rp)\okep^2\rp^4$, on a $q=1.26\cdot 10^{-5}$ planet embedded in a disc with $\rho_0 \propto r^{-3/2}$, $T\propto r^{-1}$ and $\xi=-0.85$. A small thermal diffusion coefficient $\de=10^{-7}$ was used to regularize the solution near corotation. Results for two values of the softening parameter $b$ are shown in black and red, together with a simulation with $b=0.03$ and $\de=0$ (green line). Right panel: azimuthal variation of the density at $r=0.99$ at three different times for $b=0.03$ and $\de=10^{-7}$. }
\label{figtimeevol}
\end{figure*}

Equation \eqref{eqnonlin} predicts a torque that 
is proportional to the entropy gradient of the unperturbed disc, 
and is positive for negative entropy gradients.  The torque that is reached  
is comparable in magnitude to the isothermal wave, 
or Lindblad, torque for the expected value of $b \approx H/r$ 
\citep{2002ApJ...565.1257T}. 
The torque, as given by Eq. \eqref{eqnonlin}, does not diverge for $b\rightarrow 0.$ With any separation or offset of the type described above  being neglected, setting $b=0$ in Eq. \eqref{eqnonlin} gives
\begin{equation}
\mathcal{T}_\mathrm{p}=
-\frac{4\xi \xs}{\gamma } q\Sigma_\mathrm{p}\rp^3 \op^2.
\label{eqnonlin3}
\end{equation} 
Note that in this limit the only stream parameter that enters this expression is the horseshoe width $\xs.$ The torque is then increased by a factor \noindent $\xs/( \rp b+\sqrt{\xs^2+\rp^2b^2})$~compared
 to the case $b \ne 0.$ Using Eq. (\ref{Masxs}) to estimate $\xs,$ this amounts to a factor of $4$
when $b=0.6H/r$ for $q=1.26\times 10^{-5}.$

However, its use in this limit is inappropriate. Following \cite{2006ApJ...652..730M}, we note that when vertical stratification is considered it would be expected to produce an effective softening parameter $b=z/\rp$ for material moving at height $z.$ Adopting this and performing
a vertical average of Eq. (\ref{eqnonlin}) for the above example would indicate $b \sim 0.5H/r,$
being comparable to values adopted here.

\subsection{Time development of the non-linear torque}
We studied  the time dependence of the 
disc response as a result of the  insertion of  a planet of $4$ $\me$
into a disc with $\rho_0 \propto r^{-3/2}$ and $T \propto r^{-1}.$
The model was adiabatic with $\de=0$ and  $\xi = -0.85$, in this case
corresponding to  $\gamma = 1.1.$ Note that, as mentioned in \cite{3Dpaper}, there is no smooth transition to the locally isothermal torque for $\gamma \rightarrow 1$, because in the latter case entropy is not conserved along streamlines (if there is a radial temperature gradient). In fact, for a negative density slope, a lower value of $\gamma$ tends to make the entropy gradient more negative, and therefore the associated corotation torque more positive.  

Unlike for the other simulations considered in this paper,
 in order to study the
time development of the non-linear torque in a way corresponding to the
discussion of the simple model, the planet was initially
immersed in the disc with its full mass.
Thus there was no period of slow build up for these cases.
From the streamlines illustrated in Fig. \ref{figstreamline} and discussed above
in Sect. \ref{secNonlin}, the inner and outer separatrices for a  model with $b =0.03,$  away from the planet, lie on circles
centered on the central object with dimensionless radii $0.982$ and $1.012$ respectively.

 In order to study the
development of the high-density region leading the planet  which is responsible for the
non-linear torque, we plot the density
as a function of azimuthal angle, $\varphi,$  as viewed along the radius $r=0.99$ after $5,$ $10$ and $20$ orbits in the right panel of Fig. \ref{figtimeevol}. The planet is located at $\varphi = \pi$
and for higher values of $\varphi,$ this radial location lies in the required
high density region. The left panel of the same figure shows the total torque as
a function of time (black line). A small thermal diffusivity, $\de=10^{-7}$ was used to regularize the solution around corotation. For comparison, we also show the total torque for a simulation with $\de=0$ (green line). The torque in this case rises to a slightly higher value at $t=10$, followed by a slight decay. This effect is due to the sudden introduction of the planet into the disc, and it can be reduced by introducing a small thermal diffusivity, as shown in Fig. \ref{figtimeevol}, or by slowly ramping up the planet's mass.
 
We see that as time progresses an approximately uniform
 density increase by a factor of about
$1+5\times 10^{-2}$ is established. This is entirely consistent with the factor  
 $\delta \rho/\rho = 1.55 x/r,$
with $x$ being the distance to the coorbital radius, expected from the discussion above.
The uniform density increase is also consistent with a uniform pressure
equal to the unperturbed value being attained at the fixed radius.

Let us now check that the time development is consistent
with fluid elements moving from one side of the horseshoe region to the other.
Assuming this to be the case, taking account only of Keplerian shear
 so neglecting any time required to cross close to the
planet, 
it would take $5$ orbits to set up the high density region out to
$\varphi =3.4,$ corresponding to $5$ disc scale heights from the planet, 10 orbits to set it up out to
$\varphi =3.64$ and $20$ orbits out to $\varphi=4.14.$
However, Fig. \ref{figtimeevol} shows that the density profile is not fully established
after $5$ orbits, but that after $10$ and $20$ orbits the profile is indeed
established out to the expected locations. This indicates that the evolution is relatively
slow only in the regions close to the planet. This could be related
to passage close to the separatrices where low velocities are expected.

Note too that half of the total contribution to the corotation torque we have
considered here is attained after only $5$ orbits which is  much
less than the libration period
of about $75$ orbits which is again consistent
with most of the contribution coming from close to the planet.
The magnitude of the non-linear contribution to the torque predicted by Eq. \eqref{eqnonlin} is approximately $1000$ in units of $q^2 \Sigma(\rp)\okep^2\rp^4$, which agrees perfectly with the difference between the minimum ( or also approximately the linear torque) and the maximum torque  in the left panel of Fig. \ref{figtimeevol}. This suggests that there is no additional torque cut-off close to the planet, since  one was
 not included in the non-linear model presented here. From Fig. \ref{figstreamline} can be seen that indeed the corotation feature extends to about a smoothing length from the planet in azimuth. Note too that although the associated relative density change
 is only a few percent, on account of the small radial scale the relative
 change to the density gradient is at least of order unity. Additional non-linearity
 not represented in the simple model may be present and possibly be responsible
 for the variations of the ultimate 
 ratio of torque to mass ratio squared seen in Fig. \ref{fig4}.
 The fact that these ratios are affected by small diffusion coefficients
 in the manner illustrated in Fig. \ref{fig13} indicates localized effects
 in the coorbital zone are responsible. A possibility is that because density jumps across the horseshoe separatrix are higher for larger masses, because the horseshoe region is wider, non-linear effects are also relatively stronger. 

Also shown in the left panel of Fig. \ref{figtimeevol} is the total torque for a run with a smaller value of the smoothing length $b=0.02$. From Eq. \eqref{eqnonlin2}, with $\lambda=1$, we expect the non-linear torque to be a factor 2 stronger in this case, and this is entirely consistent with Fig. \ref{figtimeevol}. We point out that from this trend the total torque would be expected to be always negative for $b \ge H/r.$

\begin{figure}
\centering
\resizebox{\hsize}{!}{\includegraphics[]{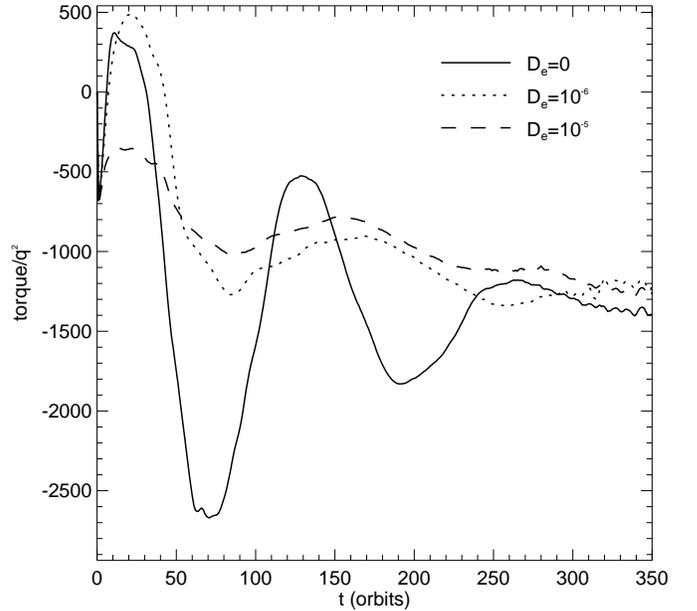}}
\caption{Long-term evolution of the total torque, in units of $q^2 \Sigma(\rp)\okep^2\rp^4$, on a $q=1.26\cdot 10^{-5}$ planet in a disc with $\rho_0 \propto r^{-3/2}$, 
$T \propto r^{-1}$ and $\xi=-0.85$, for three different thermal diffusivities.}
\label{figsaturation}
\end{figure}

\begin{figure*}
\includegraphics[width=\textwidth]{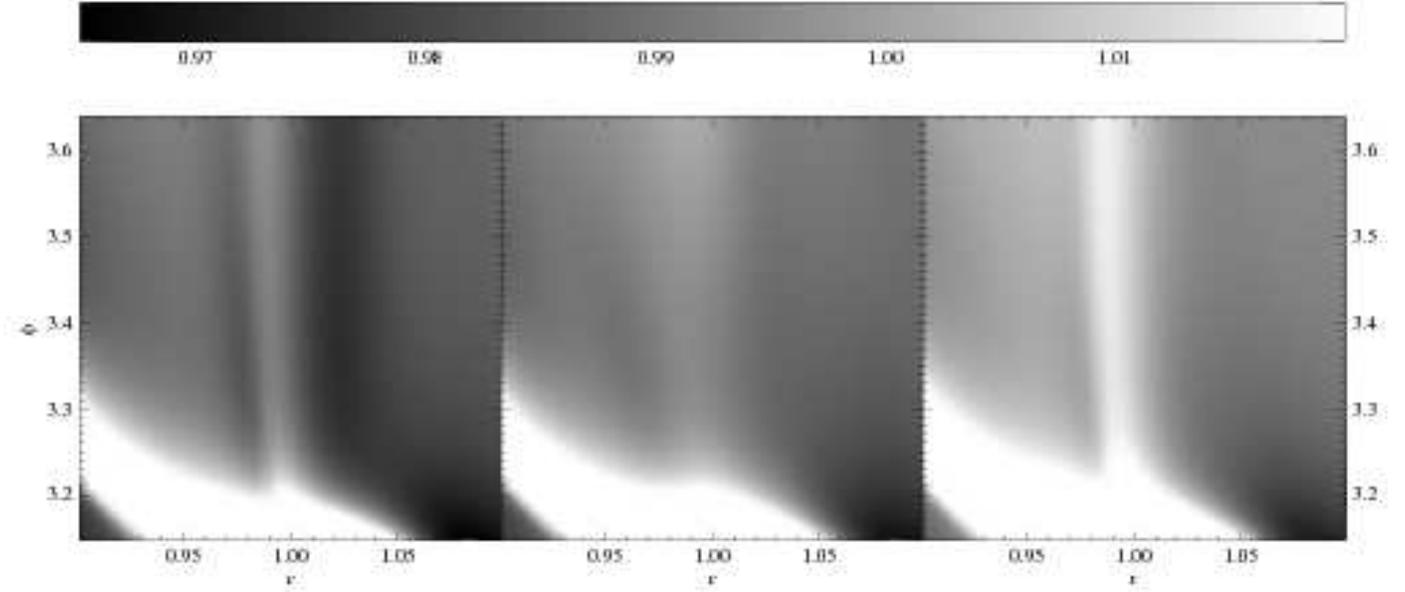}
\caption{Density, multiplied by $r^{3/2}$, for three models for $q=1.26\cdot 10^{-5}$, $\rho_0 \propto r^{-3/2}$, $T \propto r^{-1}$ and $\xi=-0.85$  after 160 orbits, focusing on the horseshoe turn leading the planet. Left panel: $\de=10^{-6}$, $\nu=0$. Middle panel: $\de=10^{-5}$, $\nu=0$. Right panel: $\de=\nu=10^{-6}$.}
\label{fig3cont}
\end{figure*}

\section{Long-term evolution}
\label{secLongterm}

We now turn to the question of the behavior of the torque over longer time scales, focusing on a disc with no vortensity gradient. We initialize the temperature $T \propto r^{-1}$, and again take  $\xi=-0.85$. As we need to follow the evolution of the torque for several libration cycles, which, for a $q=1.26\cdot 10^{-5}$, amounts to several hundred orbital periods, we lowered the resolution by a factor 2 in all directions to keep the problem computationally tractable. As mentioned in Sect. \ref{secHeatDiff}, for a finite value for the thermal diffusion coefficient the torques agreed very well for the first 30 orbits. We focus on a $q=1.26\cdot 10^{-5}$ planet throughout, and, to keep matters as simple as possible, do not ramp up the mass of the planet over the first 3 dynamical time scales as was done in the previous sections. Again, for a finite thermal diffusion coefficient, this does not influence the torques at all, while for $\de=0$ differences remain well within $10$ $\%$.

In order for a diffusive process to prevent the corotation torque from saturating, it is necessary that the diffusion time scale across the horseshoe region is smaller than the libration time scale, or:
\begin{equation}
\frac{\xs^2}{\rp^2 \op\de} < \tau_\mathrm{lib} = \frac{8}{3} \frac{\rp}{\xs}\op^{-1}.
\label{eqlib}
\end{equation}
Using our estimate Eq. \eqref{eqxs} for $\xs$, which has been shown to be a good approximation, we find for  
$q=1.26\cdot 10^{-5}$ and $b=0.03$ that $\de > 3/8(\xs / \rp)^3 \approx 3\cdot 10^{-6}$ to allow thermal diffusion the possibility of restoring the entropy gradient within one libration cycle. On the other hand, according to Fig. \ref{fig13}, for $\de=5.6\cdot 10^{-6}$ the disc response is still sufficiently non-linear that the torque is changed considerably. Therefore, there exists a range of $\de$ for which there may be a possibility of a sustained positive torque contribution from corotation effects related to the entropy gradient. However, thermal diffusion only affects the temperature gradient directly, and it may be the case that a consequent  adjustment of the density in the horseshoe region allows it to settle towards a state of zero torque contribution.

In Fig. \ref{figsaturation}, we show the long-term evolution of the total torque for three values of $\de$. The case of zero diffusion shows the clearest libration cycles, 
diminishing in amplitude until the system is expected to settle into a state of vanishing entropy gradient in the mean and therefore zero corotation torque. 
This behavior is very similar to that  described by \cite{2007LPI....38.2289W} for the barotropic case with a radial vortensity gradient.
Interestingly, both diffusive runs approach the same value of the torque, which indicates that indeed the density is redistributed in such a way that the corotation torque vanishes.
 
The final negative torque of magnitude $\sim 1200$ compares with the linear value (based on the original density distribution) of $\sim 700.$
For comparison the result of \cite{2002ApJ...565.1257T}
obtained by use of Eq. (\ref{compare})
for the original density distribution gives a magnitude of $\sim 900.$
As indicated above this was in much better agreement with the torque obtained for
the locally isothermal case with the same original density distribution.
The most likely reason for the greater discrepancy in this case is that the original density gradient in the coorbital region and its neighborhood has been modified as indicated in Fig. \ref{fig3cont} in an asymmetric manner.

In the final state, the corotation region contains an asymmetric density structure. This can be seen in the two leftmost panels of Fig. \ref{fig3cont}, where we show the density after 160 orbits for the cases with low and high diffusion. In the leftmost panel, it is clear that the low density ridge created in the horseshoe turn below the planet returns on the other side. For the higher value of $\de$, a similar situation arises, but since diffusion for this case is actually strong enough to affect the density jump directly, the effect is harder to spot. From Fig. \ref{figsaturation}, however, we see that  in this case approximately the same  total torque is  attained. 

\begin{figure}
\centering
\resizebox{\hsize}{!}{\includegraphics[]{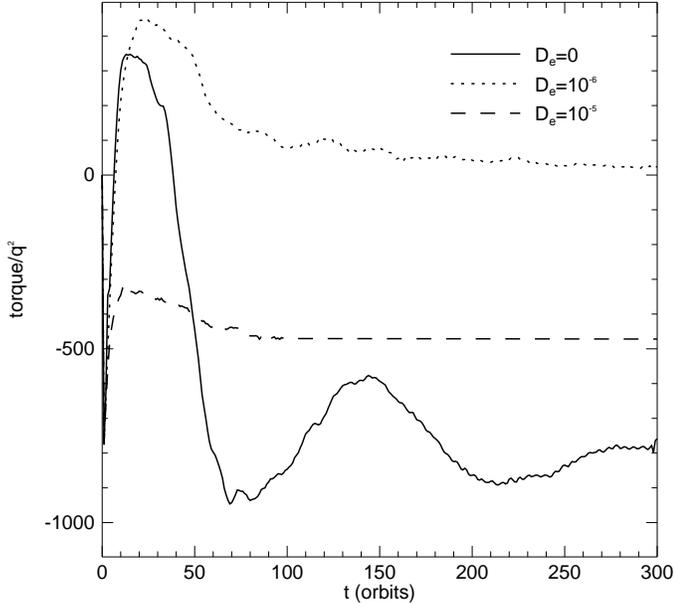}}
\caption{Long-term evolution of the total torque, in units of $q^2 \Sigma(\rp)\okep^2\rp^4$, on a $q=1.26\cdot 10^{-5}$ planet in a disc with $\rho_0 \propto r^{-3/2}$, $T \propto r^{-1}$ and $\gamma=1.1$, for three different thermal diffusivities. All models have a small kinematic viscosity of $\nu=10^{-6}$ $\rp^2\op$.}
\label{figviscsat}
\end{figure}

All of this indicates that in order to sustain the positive torque, 
 we need to restore the initial density profile as well 
as the initial temperature profile
in the coorbital region. 
This can be done by including a small kinematic viscosity. 
In Fig. \ref{figviscsat}, we show the torque evolution for the same models 
as in Fig. \ref{figsaturation}, but now including a viscosity $\nu=10^{-6}$ $\rp^2\op$. 
The mass flow induced by this small value of $\nu$ is not strong enough to
 affect the results; in essence the viscosity only acts on the sharp density features near corotation. 
We see that for a small viscosity in combination with a small thermal
 diffusivity the positive torque can be sustained. 
The magnitude strongly depends on the magnitude of the diffusion coefficients, 
as is illustrated by the dashed line in Fig. \ref{figviscsat}. 
A simulation with no heat diffusion shows similar behavior as the runs depicted 
in Fig. \ref{figsaturation}, but the final value of the torque agrees better with linear theory, presumably because of the smoother density profile. This indicates that indeed we need to restore both
 the original temperature gradient (through heat diffusion)
 and smooth the density gradient (though a small viscosity) in the coorbital region
 in order to sustain the positive torque.

These results indicate that the entropy-related corotation torque alone is indeed able to reverse the sign of the total torque over longer time scales than the libration period. Although the torque in Fig. \ref{figviscsat} is only slightly positive at later times, note that this model represents a case for which the entropy-torque alone has to overcome the full Lindblad torque. A higher value of $\gamma$, which reduces the Lindblad torque, and a flatter density profile, which gives rise to a positive corotation contribution due to the vortensity gradient, may help to push the torque to higher positive values. On the other hand increasing $\gamma$ may act to reduce the entropy gradient. The investigation of how the vortensity and entropy corotation torques operate together will be the subject of a future study.
 
\section{Discussion}
\label{secDisc}

In this paper, we have considered the torques induced by disc-planet interaction 
when low-mass planets are introduced into protoplanetary discs with  
an initial radial entropy gradient and either solved the energy equation or 
adopted an adiabatic or locally isothermal equation of state. We first considered the linear theory of the interaction. As the response is singular it needs to be regularized. This was done by adopting the well-known Landau prescription, together with a small thermal diffusivity in some cases. This  allows migration torques on the planet to be calculated, but the density response diverges as dissipative effects are reduced to zero implying that non-linear effects will ultimately be important. We estimated  that non-linear effects would be significant for dimensionless thermal diffusivities $D_e < 10^{-(5-6)}$ and that in such cases the linear theory would only be valid for some finite time after insertion of the planet. This was confirmed by the non-linear simulations. It was found from these that a non-linear density structure developed around corotation at early times, that in some cases with a negative entropy gradient, led to a reversal of the sign of the torque leading to outward migration.
This positive torque survives for a horseshoe libration period after which saturation occurs and the torque again becomes negative.

We presented a simple non-linear model to account for the non-linear coorbital density structures as being produced after the  horseshoe turns undergone by material while conserving its entropy in Sect. \ref{secNonlin}. Numerically we found that in cases with no diffusivity, the coorbital density structure showed variations on the grid scale and therefore in order to get numerical convergence at limited resolution, a small non-zero thermal diffusivity needs to be included. Increasing the thermal  diffusivity brings the migration torque closer to the linear prediction although some nonlinearity associated
with the torque connected with the specific vorticity gradient may remain.

The non-linear and temporary nature of the physical effect that causes the torque to become positive makes it inappropriate to attempt to deduce an applicable  linear torque formula, as was done in the barotropic case \citep{1979ApJ...233..857G}. All that can be said is that this non-linear part of the torque depends on the entropy gradient. We have found that, for a disc with $H/r=0.05$, a density gradient less steep than $\rho_0 \propto r^{-3/2}$ results in a positive torque on the planet. However, this depends also on the adopted value of the smoothing length, 
 the value of $\gamma$ and the amount of thermal diffusion. A smaller value for $b$ will enhance such non-linear effects, while thermal diffusion acts in the opposite direction. 

Our simple non-linear model of Sect. \ref{secNonlin} does a good job in qualitatively describing the results of the numerical simulations. Because the three-dimensional simulations of \cite{2006A&A...459L..17P} show the same slow rise of the torque, which is indicative of non-linear effects, it is likely that a similar model applies in 3D.

The strong dependence of low-mass planet migration on the radial entropy gradient in the gas disc raises two important questions. First of all: do realistic, self-consistent disc models with detailed radiative transfer allow for negative entropy gradients? Second: is the thermal diffusivity low enough to support non-linearity at corotation? The second question was already answered by the simulations of \cite{2006A&A...459L..17P}, which showed that, at least in the inner parts of protoplanetary discs, radiative transport does not provide enough diffusion to restore linearity. As opacity  in general decreases with radius, the situation will be different in the outer parts of the disc.

The first question is more difficult to answer. Detailed models of \cite{1998ApJ...500..411D,1999ApJ...527..893D,2001ApJ...553..321D} show a rich variety in temperature and density profiles, with indeed several cases with strongly negative entropy gradients, especially near opacity jumps. These may be preferential radii for halting planet migration, a conclusion also reached by \cite{2004ApJ...606..520M} for different reasons. Although it is beyond the scope of this paper to investigate the occurrence of negative entropy gradients in detail, it is important to realize that they do appear in detailed models.

In the inviscid limit, the corotation region is a closed system,
 and can therefore not continue to exert a torque on the planet indefinitely. 
The first signs of saturation of 
the torque are already visible in,
 for example, Fig. \ref{fig2} and the  complete process is illustrated in Fig. \ref{figsaturation}.
 In order to sustain this torque for longer
times, the physical conditions in the coorbital region need to 
be regenerated. Heat diffusion can act to restore the original entropy gradient, but we have found that  the form of the  surface density gradient also  needs to be smoothed
locally, which can be achieved by including a small kinematic viscosity.

At this point we stress the sensitivity of torque behavior derived
from  effects at corotation to small values of thermal diffusivity and
viscosity. This has already been indicated by
\cite{2002A&A...387..605M} who found that small values of viscosity
acting in conjunction with effects at corotation could have
significant effects  on the migration of planets but in a larger mass
range than that considered here.  
 

\section{Summary and conclusion}
\label{secCon}
In this paper, we have studied the effect of including the energy
equation in the theory of disc planet interactions undergone by
low-mass planets. We have shown that the wave torque on the planet is
reduced by a factor $\gamma$ compared to locally isothermal discs. It
is the corotation torque, however, that is the most strikingly
different from the isothermal case. 

We have found that migration behavior depends critically on the radial
entropy gradient in the disc. The linear estimate of the corotation
torque depends on this, but the entropy gradient is also responsible
for a departure from linearity, giving rise to a strong corotation
torque that is positive for a negative entropy gradient. For a strong
enough entropy gradient these corotation torques can dominate over the
wave torques. This can be understood in a qualitative way by
considering simplified horseshoe dynamics. Thermal diffusion reduces
non-linearity, and for high enough values of the thermal diffusion
coefficient $\de$ we recover  modified  linear wave torques and a
mildly non-linear barotropic corotation torque.  

Although for a planet on a fixed orbit and with no viscosity, the
positive torques saturate after a libration period, we have shown
that, for a small thermal diffusivity and viscosity the
entropy-related corotation torque offers the possibility of a
sustained reversal of the direction or stalling of migration of
low-mass protoplanets. Of course extensive future investigations that
take into account three dimensional effects and allow the planet to
migrate are required in order to evaluate it. 

\bibliographystyle{aa} 
\bibliography{8702.bib}

\end{document}